\documentclass[useAMS,usenatbib]{mn2e}
\usepackage{subcaption}
\usepackage{ifpdf}

\ifpdf
\usepackage[pdftex]{graphicx}
\else
\usepackage{graphicx}
\fi

\ifpdf
\DeclareGraphicsExtensions{.pdf, .jpg, .tif, .png}
\else
\DeclareGraphicsExtensions{.eps, .jpg}
\fi

\title[Globular Cluster Formation in the Virgo Cluster]{Globular Cluster Formation in the Virgo Cluster}
\author[C. Corbett Moran, R. Teyssier, G. Lake ]{C. Corbett Moran$^{1}$\thanks{E-mail:
corbett@physik.uzh.ch}, R. Teyssier$^{1}$, G. Lake$^{1}$\\
$^{1}$Institute for Theoretical Physics, University of Zurich, Winterthurerstrasse 190, CH-8057, Z\"urich Switzerland}
\begin{document}
\date{Received 2013 December 10}

\pagerange{\pageref{firstpage}--\pageref{lastpage}} \pubyear{2013}

\maketitle

\label{firstpage}

\begin{abstract}
Metal poor globular clusters (MPGCs) are a unique probe of the early universe, in particular the reionization era. Systems of globular clusters in galaxy clusters are particularly interesting as it is in the progenitors of galaxy clusters that the earliest reionizing sources first formed. Although the exact physical origin of globular clusters is still debated, it is generally admitted that globular clusters form in early, rare dark matter peaks \citep{Moore:2006bi,Boley:2009gf}. We provide a fully numerical analysis of the Virgo cluster globular cluster system by identifying the present day globular cluster system with exactly such early, rare dark matter peaks. A  popular hypothesis is that that the observed truncation of blue metal poor globular cluster formation is due to reionization \citep{Spitler:2012fr,Boley:2009gf,Brodie:ip}; adopting this view, constraining the formation epoch of MPGCs provides a complementary constraint on the epoch of reionization. By analyzing both the line of sight velocity dispersion and the surface density distribution of the present day distribution we are able to constrain the redshift and mass of the dark matter peaks. We find and quantify a dependence on the chosen line of sight of these quantities, whose strength varies with redshift, and coupled with star formation efficiency arguments find a best fitting formation mass and redshift of $\simeq 5 \times 10^8 \rm{M}_\odot$ and $z\simeq 9$.  We predict $\simeq 300$ intracluster MPGCs in the Virgo cluster. Our results confirm the techniques pioneered by \cite{Moore:2006bi} when applied to the the Virgo cluster and extend and refine the analytic results of \cite{Spitler:2012fr} numerically.
\end{abstract}

\section[Introduction]{Introduction}

\label{sec:introduction}

Cosmic reionization of the intergalactic medium, in which its hydrogen gas changes from a neutral to an ionized state, is fundamentally intertwined with galaxy formation at high redshift. After recombination at $z\sim1000$ \citep{2013ApJS..208...19H} gas in the universe was primarily neutral, but observations of the Gunn-Peterson trough in quasar spectra indicate that the universe was fully reionized by $z\sim6$ \citep{1965ApJ...142.1633G,1965Natur.207..963S}. Further observations, from the evolution of the $L\alpha$ galaxy luminosity function \citep{Nagamine:2010tt} to the CMB polarization anisotropy \citep{2011ApJS..192...16L} indicate that this process was not universally instantaneous and extended from as early as $z\sim14$ to $z\sim6$ \citep{Fan:2006he}. In general, ionizing sources appeared first in high density environments \citep{2009ApJ...703L.167A}

To probe the reionization epoch, one method is to use the thousands of Globular Clusters (hereafter GCs) with masses from $10^4$ to $10^6$ solar masses which are detected in the Virgo Cluster \citep{Peng:2008bp}
or in other nearby galaxies \citep{Rhode:2005jv} at the present day. 

GC color distribution is observed to be almost universally bimodal \citep{1985ApJ...293..424Z,Gnedin:2009tya,EricHNeilsen:gh} with the blue, metal poor GCs (MPGCs) being older and having radial distributions and chemical compositions similar to halo field stars \citep{2008A&ARv..15..145H}. This suggests two formation channels for the two populations \citep{Brodie:ip} with a physical mechanism separating the formation of the two groups by the observed spread of $\sim 1.5 \rm{Gyr}$. MPGCs are of primary importance because they host the oldest stars in the whole cluster population and probe conditions at high redshift, during the earliest stages of galaxy formation \citep{Chaboyer:303494}, exactly the period in which cosmic reionization is taking place. Thus MPGCs provide what amounts to a fossil record of conditions during their formation billions of years ago.  

A consistent picture has emerged where MPGCs form in rare overdense fluctuations and invokes reionization to truncate their formation successfully explaining the bimodal population \citep{Forbes:2006uv}. Overdense fluctuations are the first to collapse in a cosmological context \citep{1999MNRAS.308..119S} and thus are expected to share the same spatial distribution as old stellar populations exactly such as metal poor globular clusters and halo field stars at the present day. Associating the protogalactic fragments of Searle {\&} Zinn \citep{1978ApJ...225..357S} with the rare peaks able to cool gas in the CDM density field at $z>10$ this indicates that metal poor globular clusters formed when the universe was less than a billion years old \citep{Moore:2006bi}. Such a formation period is consistent with formation of the population before reionization  and lends itself naturally to the hypothesis that the cold molecular clouds within globular clusters are heated by the reionization process to the point star formation is no longer favored \citep{Forbes:2006uv} truncating globular cluster formation temporarily. After some time in this picture, the second metal rich population begins to form in galaxy merger events, when the gas within the host galaxies has had the chance to enrich \citep{Brodie:ip}.

\cite{Muratov:2010gq} present an alternate scenario in which MPGCs form in merger events, which at later times could produce both metal poor and metal rich GCs. Current measurements of cluster ages in our Galaxy can neither refute the reionization truncated scenario nor support the merger based, thus more extended in time, scenario of MPGC formation favored by \cite{Muratov:2010gq} as these surveys cannot detect age differences below $\sim 1$~Gyr \citep{Dotter:2009ek,MarinFranch:2009fb,DeAngeli:2005en}. A narrow age dispersion, predicted by the reionization truncated formation scenario, remains consistent with the data \citep{DeAngeli:2005en,MarinFranch:2009fb}.

In this paper, we adopt theoretical theoretical scenario in which MPGC formation is truncated by reionization focusing on a Virgo Cluster like halo allowing us to explore its consistency by comparing with M87 observational data. M87 is especially interesting theoretically as it resides in the high density Virgo Cluster environment and hosts the most populous globular system of nearby galaxies \citep{Cote:2001bl}, providing a rich bed of observations to compare to \citep{Hanes:2001hj,Cote:2001bl}. We explore the consequences of the reionization truncated formation scenario, providing hints as to the reionization epoch in a cluster environment and constraints on alternate formation scenarios.

The rare density peaks at high redshift that MPGCs form in later merge into larger objects, namely large  galaxies like the Milky Way \citep{1994MNRAS.271..676L,1974ApJ...187..425P}. A large fraction of these galaxies ultimately end up in massive clusters of galaxies today. This three stage hierarchical process is very difficult to capture unless one uses ultra high-resolution simulations. Such simulations give us to opportunity to explore this formation scenario by affording us the ability to trace such peaks matching them with  present day globular cluster systems and following their kinematic and spatial distribution properties from formation until the present day.

In this paper we are able to push our resolution capabilities to the requisite level by capturing MPGC evolution using a two stage numerical procedure. First, using ultra high resolution N-body simulations in the cluster region, we followed dark matter halos down to $z\sim 4$, where the old MPGC population must be in place \citep{Moore:2006bi,Boley:2009gf}. We then went back to our cluster sample and used our high-redshift MPGC catalog to mark dark matter particles as potential MPGC tracers. A halo mass cut M chosen at given redshift $z$ corresponds to the peak overdensity $\nu$, with higher masses at higher redshifts corresponding to more over dense, or higher-$\nu$ peaks. Only particles above a requisite $\nu$ are marked. Since MPGCs are very dense and bound objects, once $\nu$ is fixed we are able to predict the MPGC population in satellite and central galaxies as well as ``free floating'' MPGCs. An advantage of this procedure with respect to simulation physics, known hereafter as the Diemand-Moore technique, is that it is independent of the detailed baryon physics \citep{Boley:2009gf}. By using a suite of reference simulations at multiple resolutions, in which both higher and lower resolution simulations are able to be performed down to $z=0$, we are able to show that this matching procedure is indeed effective and preserves the relevant physical properties we use to compare to observations, namely the globular cluster system surface density profile and the line of sight velocity dispersion profile.

The Diemand-Moore technique was first shown to accurately reproduce the present day globular cluster system in the Galaxy \citep{Moore:2006bi} via an ultra high resolution N-body simulation. \cite{Spitler:2012fr} use this technique to provide evidence for inhomogeneous reionization in the local universe from MPGCs. The Diemand-Moore technique was introduced to use the spatial concentration of MPGCs relative to the host galaxy's halo mass profile to constrain the rarity of the halos that the MPGCs formed within. This technique requires MPGC surface density profiles from observations, a halo mass model of the galaxy hosting the MPGCs and a framework to interpret the spatial bias of the MPGCs. Their fitting to the observed globular cluster system properties of M87, NGC 1407, and our Galaxy enabled them to derive a joint constraint on $z_{\rm{reion}}$ of $10.5^{+1.0}_{-0.9}$ with individual constraints of  $z_{\rm{reion}}=12.1^{+1.6}_{-1.1}$, $z_{\rm{reion}}=11.0^{+1.7}_{-1.7}$, $z_{\rm{reion}}=7.8^{+2.0}_{-1.8}$ for the Milky Way, NGC 1407 and M87 respectively. Intriguingly they find that while the reionization process is thought to begin earliest in high density environments, such as the Virgo Cluster in which M87 resides, it completes last in the cluster environment.

Motivated to explore and validate the assumptions behind this counterintuitive result reached by \cite{Spitler:2012fr} we continue this line of research and explore ways to relax some of the assumptions made by \cite{Spitler:2012fr}. Specifically we get rid of the the assumption of spherical symmetry, performing the analysis numerically vs. analytically, and assume molecular in addition to atomic cooling in star formation arguments used to check the consistency of our candidate redshift and mass pairs which correspond to a given density peak height, the number of standard deviations above the mean mass-density level at that epoch \citep{1994MNRAS.271..676L}. Using our simulations coupled with the matching technique, we present comparisons to observed properties in M87 and show numerical refinements to the previous analytic results and constraints of \cite{Spitler:2012fr} Under these set of relaxed assumptions and refined methods, we find a value of $z\sim9$ for the truncation redshift of MPGC formation, and identify this in our model as $z_{\rm{reion}}\simeq 9$ for the Virgo Cluster.
  
In Section \ref{sec:simulation} we describe the numerical methods adopted for our simulations in Section \ref{sec:matching} we describe our matching technique, and show that such resolution bridging is indeed possible preserving the relevant physics. In Section \ref{sec:results} we show the present day particles corresponding to high-sigma peaks at high redshift indeed reproduce the blue globular cluster distribution at $z=0$ in a  Virgo like object by comparing to the observational properties of M87. In Section \ref{sec:synthesis} we present the best fitting model, with error bars based on line of sight dependency, with associated discussion.

\section{Simulations}
\label{sec:simulation}

Using the RAMSES code \citep{Teyssier:2002fj} with initial conditions computed using the Eisenstein \& Hut transfer function \citep{1998ApJ...498..137E} computed using the Grafic++ code \citep{Potter:bWn-q5m9} we performed a suite of dark matter only zoom cosmological simulations in a $\Lambda CDM$ cosmology using with cosmological parameters set using WMAP5 results as listed in \ref{cosmologyparamstable}.

\begin{table}
 \centering
 \begin{minipage}{140mm}

\caption{Cosmological parameters for our simulations.}
\label{cosmologyparamstable}
  \begin{tabular}{ c | c | c| c | c | c}
    \hline
     \textit{$H_0$ [\rm{km} $\rm{s}^{-1} \rm{Mpc}^{-1}$]} & \textit{$\sigma_8$} & \textit{$n_s$} & \textit{$\Omega_\Lambda$} & \textit{$\Omega_m$} & \textit{$\Omega_b$} \\
    \hline
    70.4 & 0.809 & 0.809 & 0.728 & 0.272 & - \\
    \hline
  \end{tabular}
  \end{minipage}
\end{table}

The zoom technique selects a subregion of the computational domain to achieve the requisite resolution. To select this subregion, which we desire to form a halo of Virgo like mass by $z=0$ we first performed a low resolution simulation and identify dark matter halos and subhalos using the AdaptaHOP algorithm \citep{Aubert:2004im} in conjunction with a merger tree identification algorithm, as specified in \citep{Tweed:2009fh} as part of the GalICS pipeline as HaloMaker \citep{Tweed:2006wk} and TreeMaker respectively \citep{Tweed:2006wo}. Using the structure identify form a catalog of dark matter halos at each simulation output redshift and selected a subsample of halos with a mass of $1-3 \times 10^{14} \rm{M}_\odot$, on the order the observed Virgo cluster mass at $z=0$. To select our best Virgo candidate we examined assembly history using the merger trees and selected the halo that had a relatively quiescent merger history. We can consider our selected halo relaxed as its last major merger occurs at $z\sim 1.5$. The virial mass of our selected halo, which we define as $\rm{M}_{200 \rho_c}$ is $1.31 \times 10^{14} \rm{M}_\odot$ and the virial radius of our selected halo which we likewise define as $r_{200 \rho_c}$ is $1.06\rm{Mpc}$

 This halo was then re-simulated in a zoom simulation focusing the computational resources in the region in which it forms. We performed three simulations of successively higher resolutions: low, high, and ultra-high of which the mass and spatial resolution is detailed in \textbf{Table \ref{massandspatialresolutiontable}}. Only the low and high resolution simulations were run to $z=0$ and were used to validate our matching technique. This matching technique allows us to run the ultra-high resolution simulation only until $z\sim4$. Using the match technique together with the high resolution simulation were are able to analyze robustly physical quantities such as the surface density and velocity dispersion profiles at $z=0$ despite not running the ultra-high resolution simulation to the present day.

\begin{table}
 \centering
 \begin{minipage}{140mm}
\caption{Mass and spatial resolution for our simulations}
\label{massandspatialresolutiontable}
  \begin{tabular}{ l | c | c}
    \hline
    Name & $\mathit{m_{\rm{cdm}}} [10^6 \rm{M}_\odot]$ & $\mathit{\Delta x_{\rm{min}}} \rm[kpc/h]$ \\ 
    \hline
    Low & 54 & 1.5 \\
    High & 5.4 & 0.8 \\
    Ultra high & 0.54 & 0.4 \\
    \hline
  \end{tabular}
  \end{minipage}
\end{table}

\section{Halo Matching Procedure}
\label{sec:matching}
Steps in the matching procedure are detailed as follows:
\begin{enumerate}
  \item Run halo finder HOP \citep{1998ApJ...498..137E}, on the high resolution simulation at a given redshift. This outputs a group ID associated with each particle, which is 0 if the particle is not associated with a group.
  \item For halos above a given mass cut and below a given contamination fraction (the fraction of mass of particles at the highest resolution relative to total mass), mark the particles  within these halos. This was done via a modified version of the $\tt{poshalo.f90}$ routine available in the standard RAMSES distribution called $\tt{markparticles.f90}$. The routine $\tt{poshalo.f90}$ reads in the halos computed by HOP and outputs the number of particles, contamination factor, center and velocity of each halo. $\tt{markparticles.f90}$ is identical to this routine but additionally checks if a particle's assigned group's mass is above the mass cut and its group's contamination fraction is below the contamination cut. For the particles whose respective groups meet this criterion their IDs are written to an ASCII file on disk.
  \item Trace marked particles in the high res simulation at $z=0$ to the low resolution initial condition file at $z=105$. This was done using the $\tt{geticmask.f90}$ routine available in the standard RAMSES distribution. This is a utility that for the purpose of zoom simulations allows the user to select a group of particles-specified by their associated global particles ID in an ASCII file $\tt{partID.dat}$, typically within the same halo, and trace back to their Lagrangian volume to a Grafic initial condition file. The first RAMSES simulation output, the particle IDs to be traced, and the Grafic initial condition files are given as input. Here, as we had RAMSES particle IDs we wanted to trace back to initial conditions as an output of the previous step we could simply re-purpose the code for this, even as these particle IDs correspond to different halos with the caveat as we wanted to mark these and only these particles, we removed the Zeldovich approximation drifting so as to not mark drifted particles. This outputs a file called $\tt{ic\_refmap}$ indicating those particles ``marked'' in the Grafic initial condition file.
  \item Match the low resolution initial condition file at $z\approx 100$ to the first output of the simulation, identifying marked particles in the low resolution simulation. This was done via the $\tt{track\_particles}$ C++ routine developed as part of this work built upon Oliver Hahn's C++ library to access RAMSES snapshot files, libRAMSES++ \citep{Hahn:2008tb} and the $\tt{fio}$ library \cite{Potter:bWn-q5m9} which handles Grafic initial condition files. $\tt{track\_particles}$ uses the C++ library to read in the ramses file checks for each particle in the RAMSES simulation the library reads in if the particle lies in the inner zoom simulation box, if it does it computes its grid cell in the Grafic IC accessed using the $\tt{fio}$ library, goes to that grid cell in the the Grafic IC, check in the $\tt{ic\_refmap}$ if it is marked, and if so write the particle ID in the lower resolution simulation to a binary array of integers, with a header indicating the total number of particles.
  \item Compute the profile of the marked particles at $z=0$ in the low resolution simulation about these particles. This was done via a modified version of $\tt{cylpart2prof.f90}$ available as part of the standard RAMSES distribution called appropriately $\tt{markedcylpart2prof.f90}$, which is identical except that it additionally reads in the binary file of marked particles computed in the previous step and only profiles those particles which are marked. Thus the code takes in both a RAMSES output (we used that of $z=0$) and a marked particle file in the previously specified binary format. This gives us the desired profiled quantities we are interested in: cylindrical surface density distribution and circular velocity dispersion properties, for comparison with observation.
\end{enumerate}

\begin{figure}
  \centering
\caption{Steps in the matching procedure}
\label{matchingschematic}
  \includegraphics[width=0.95\linewidth]{{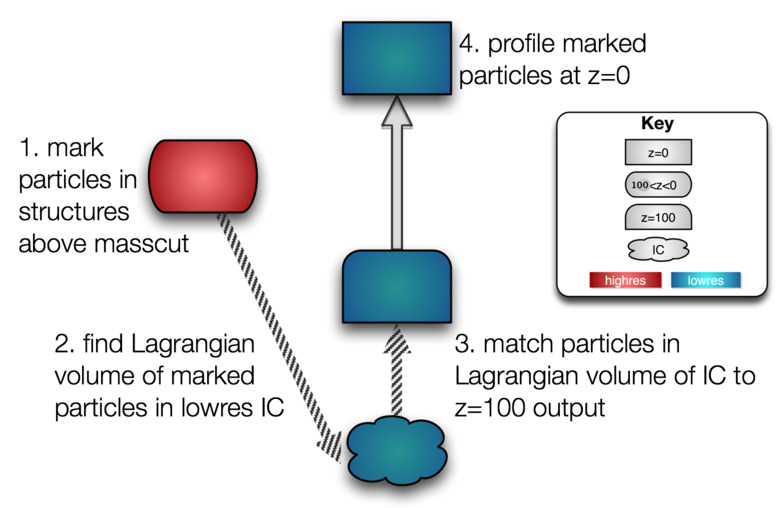}}

\end{figure}

A schematic of the matching procedure is depicted in \textbf{Figure \ref{matchingschematic}}

\subsection{Matching Technique Effectiveness} 
To verify the matching procedure we used the low and the high resolution simulations given in \textbf{Table \ref{massandspatialresolutiontable}} each of which we were able to run to $z=0$. Thus we could compare the result of the procedure above, specifically comparing the following two cylindrical surface density distribution and circular velocity dispersion properties 
\begin{enumerate}
\item[a)] the profiled quantities of low resolution particles in the low resolution simulation at $z=0$ which were marked at high redshift in the high resolution simulation.
\item[b)] the profiled quantities of high resolution particles in the high resolution simulation at $z=0$ which were marked at high redshift in the high resolution simulation.
\end{enumerate}
If a) and b) are sufficiently close we can be convinced of the technique's effectiveness when applied to the high and ultra-high resolution simulations, allowing us to forgo the cost of running the ultra-high resolution simulation to $z=0$.

\begin{figure}
\centering
\caption{Showing that the matching technique between particles marked at higher resolution to particles at low resolution is suitable for recovering profile information at lower resolution.}
\label{matchingtechnique}

\begin{subfigure}{0.95\linewidth}
  \centering
  \includegraphics[width=0.95\linewidth]{{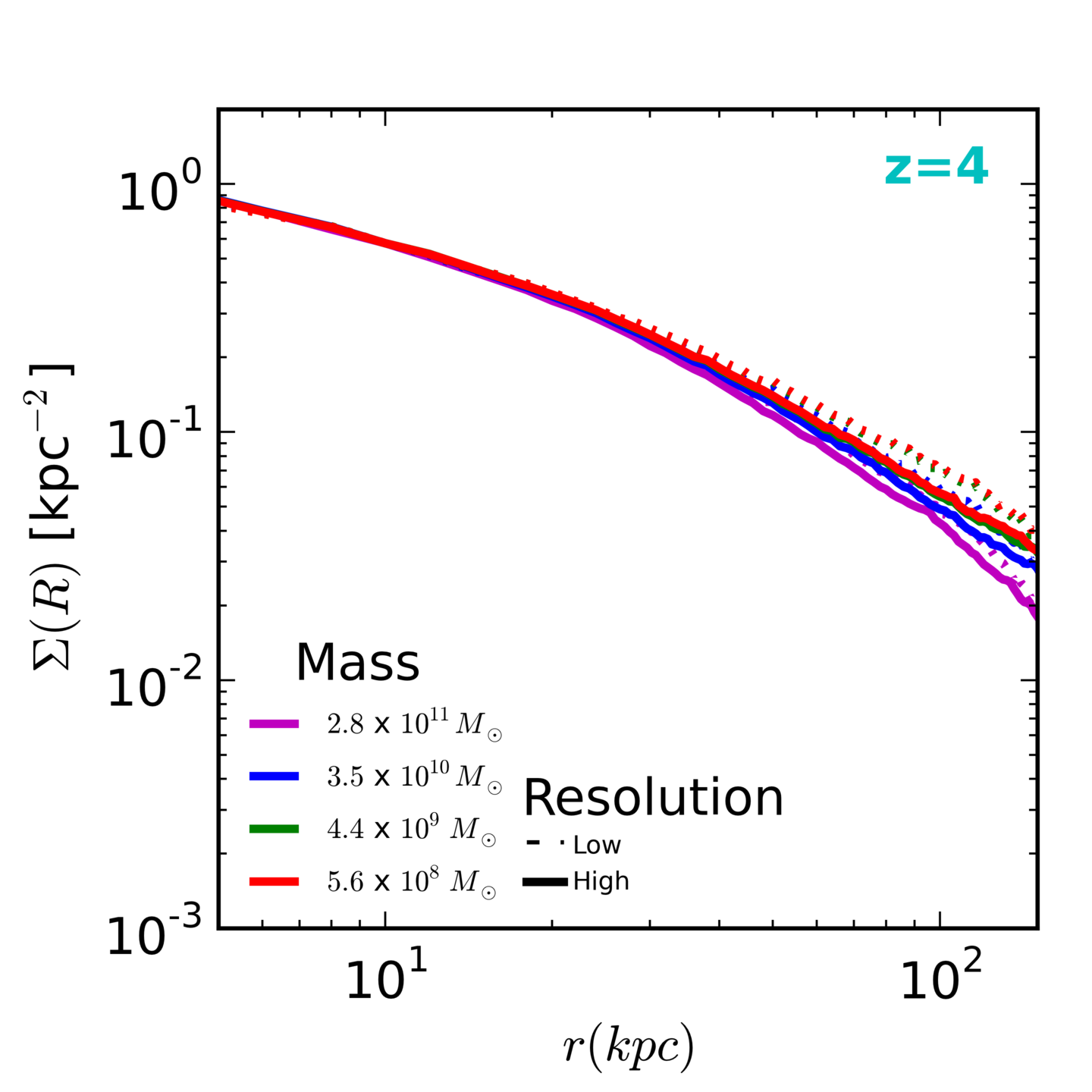}}
\end{subfigure}
\begin{subfigure}{0.95\linewidth}
  \centering
  \includegraphics[width=0.95\linewidth]{{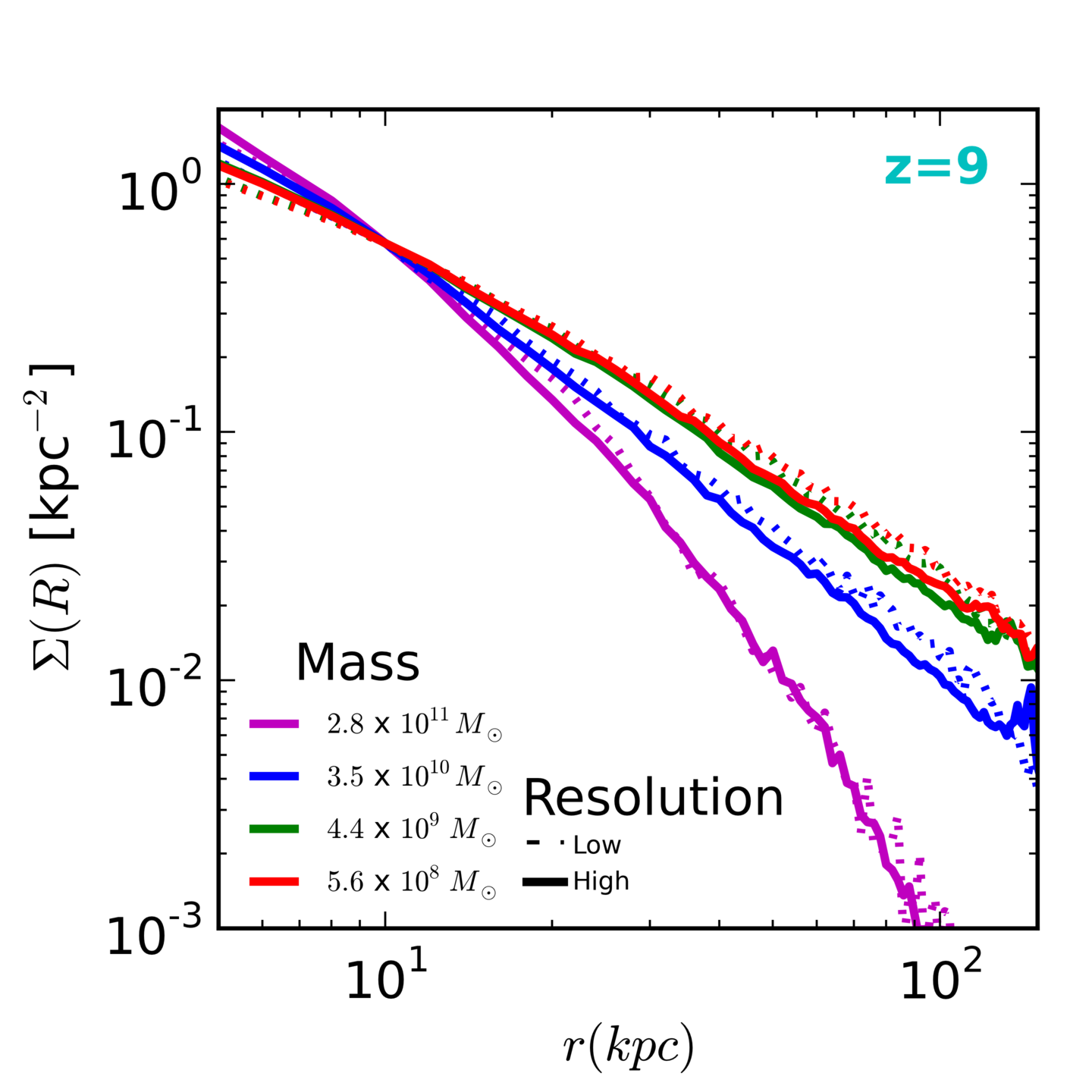}}
\end{subfigure}
\end{figure}

\textbf{Figure \ref{matchingtechnique}} shows firstly than the curves at subsequently higher resolution are consistent, with similar features, and in the case of density near identical slopes. Note that the a lower resolution particle is about 10x more massive than a higher resolution particle, so selecting 1000 particles at higher resolution, corresponds to selecting 100 particles at lower resolution. This shows that we can successfully mark particles at high redshift and higher resolution, and step down one order of magnitude in the simulation that is run to $z=0$ and still recover the correct profile for those marked particles at $z=0$, with the added advantage of having a much more detailed trace of their kinematical properties (by an order of magnitude) in the high resolution simulation at high redshift. Finally, we can see that there remains an advantage to working at the highest resolution possible, as the lower resolution profiles tend to overstate the mass, velocity, and densities, with these profiles becoming more refined at higher resolution, as would be expected.

\section{Results}
\label{sec:results}
To compare to observational data existing for M87's globular cluster system, which exists for both cylindrical surface density distribution and line of sight velocity dispersion, we computed these quantities for our simulations for various formation scenarios. The scenarios that correspond best to the observational data we deem most likely. Each of these quantities has a dependence on chosen line of sight for the measurement. To quantify this dependency, for all of our analysis we randomized over the line of sight and error bars indicate the scatter in the measured quantity over 100 such random lines of sight and the results are detailed in this section.

\subsection{Surface Density Distribution and Line of Sight Velocity Dispersion}
\begin{figure}
\caption{Variation of surface density at $z=0$ of the globular cluster population with redshift cut for a fixed mass cut. Cyan points, with associated error bars, are observational data for M87 MPGCs \protect\citep{2011ApJS..197...33S}.}
\label{sdfixedm}
\begin{subfigure}{.95\linewidth}
\centering
\includegraphics[width=.95\linewidth]{{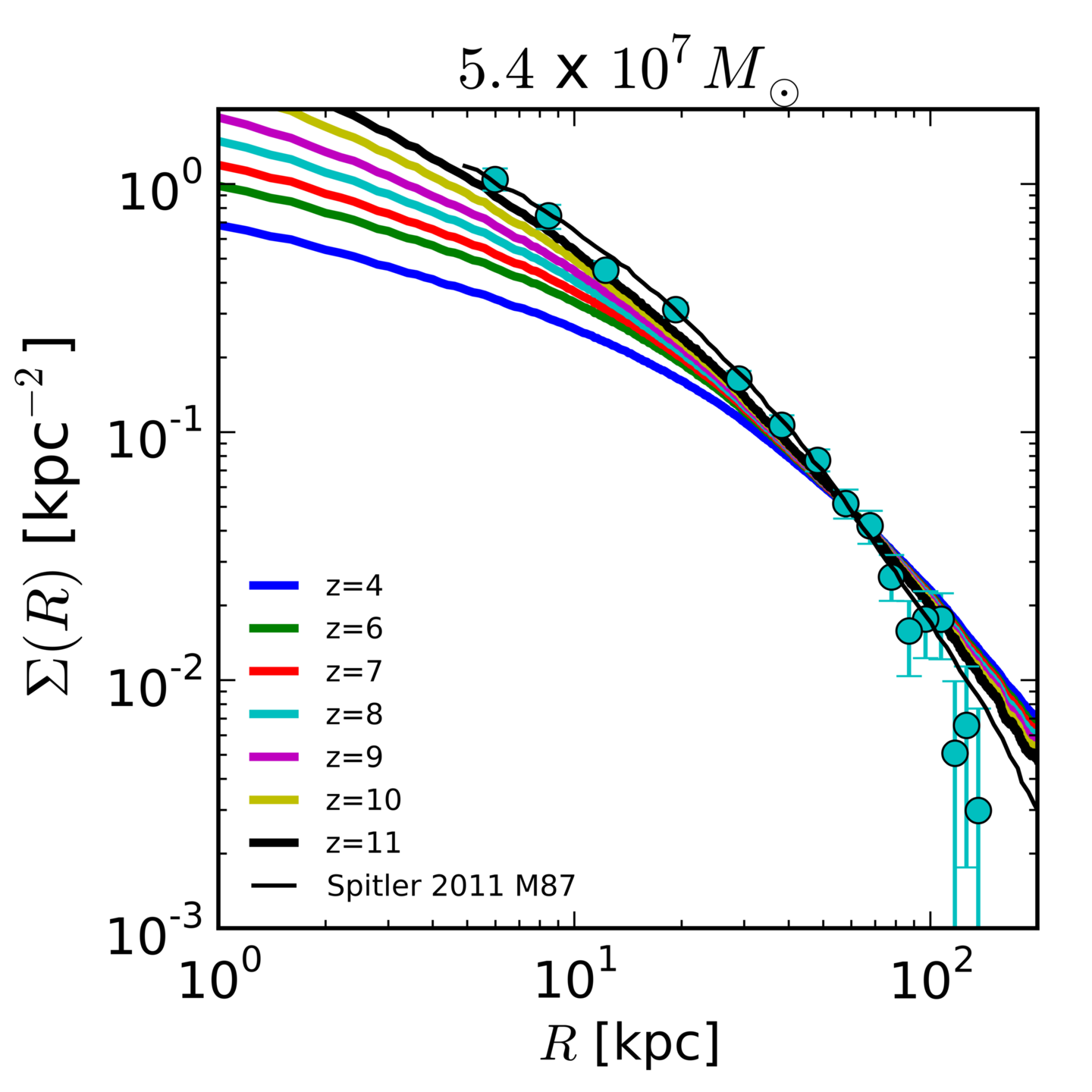}}
\end{subfigure}
\begin{subfigure}{.95\linewidth}
\centering
\includegraphics[width=.95\linewidth]{{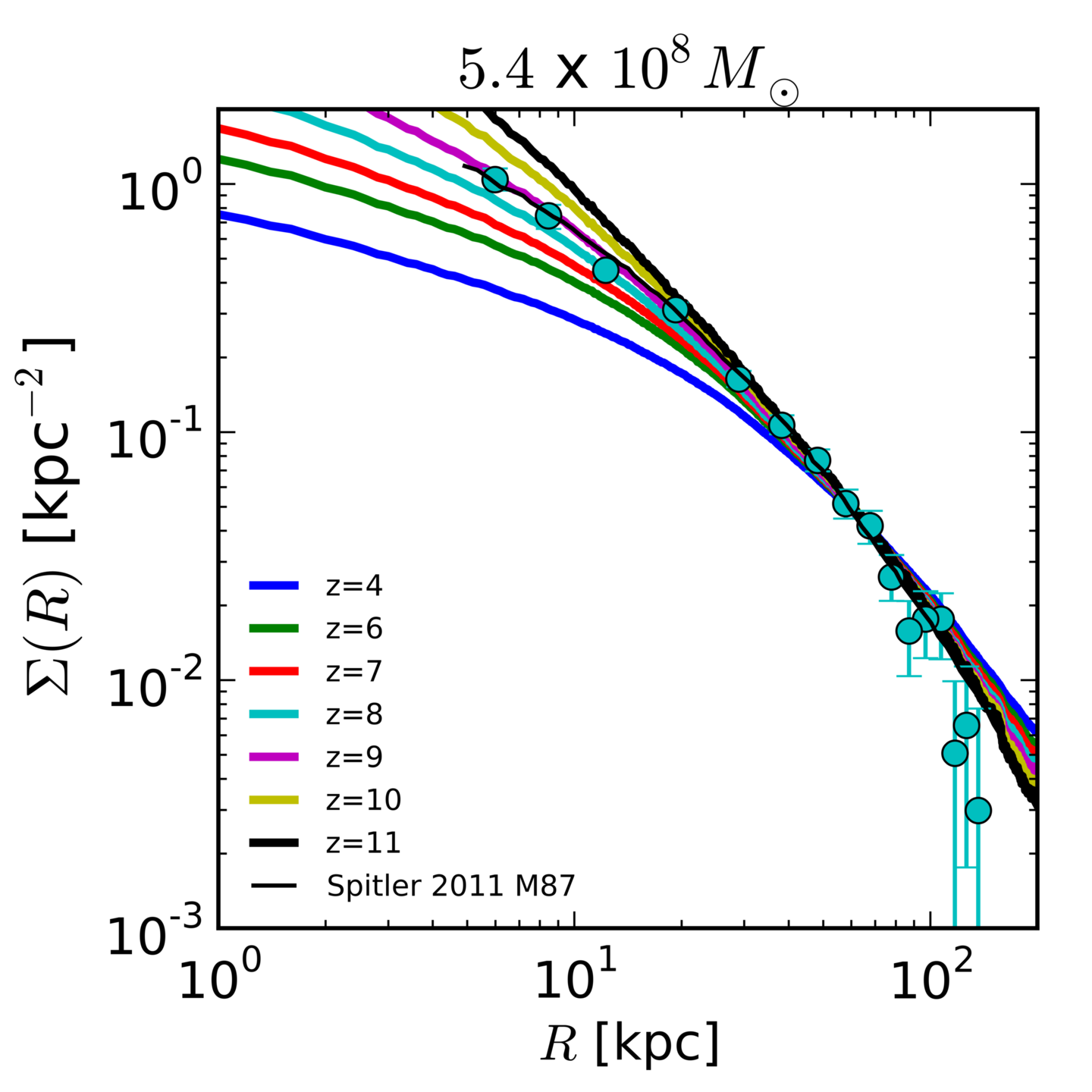}}
\end{subfigure}
\begin{subfigure}{.95\linewidth}
\centering
\includegraphics[width=.95\linewidth]{{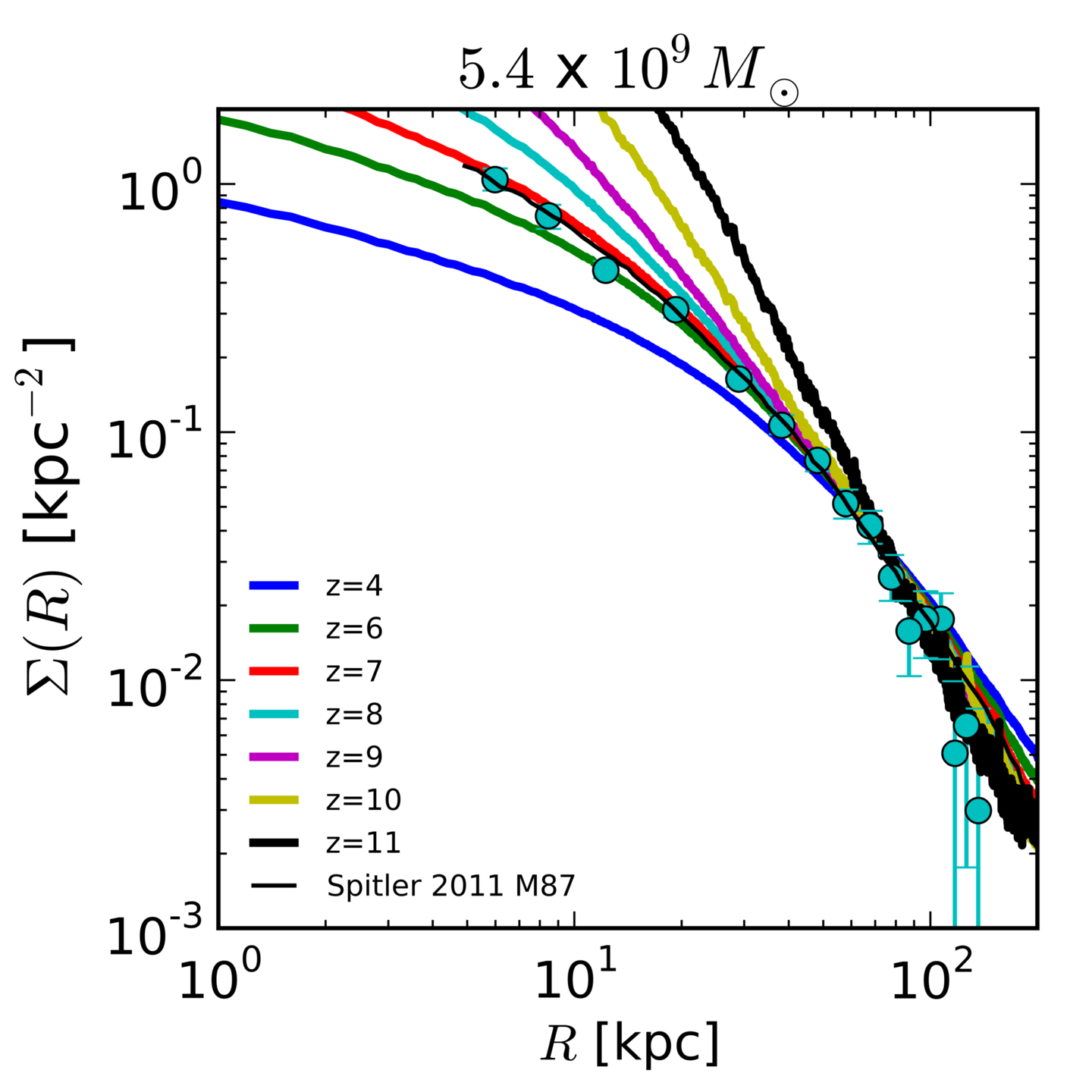}}
\end{subfigure}

\end{figure}

The variation of surface density with redshift cut for a fixed mass cut is depicted in \textbf{Figure \ref{sdfixedm}} and shows that for a mass $M\simeq 5 \times 10^7 \rm{M}_\odot$ the best fit redshift cut is $z\simeq11$ , for $M\simeq 5 \times 10^8 \rm{M}_\odot$ $z\simeq9$, for $M\simeq 5 \times 10^9 \rm{M}_\odot$ $z\simeq7$. Thus, for surface density profiles the higher the mass cut the lower the redshift cut that is consistent with the Virgo data. 

\begin{figure}
\begin{subfigure}{.95\linewidth}
  \centering
  \includegraphics[width=.95\linewidth]{{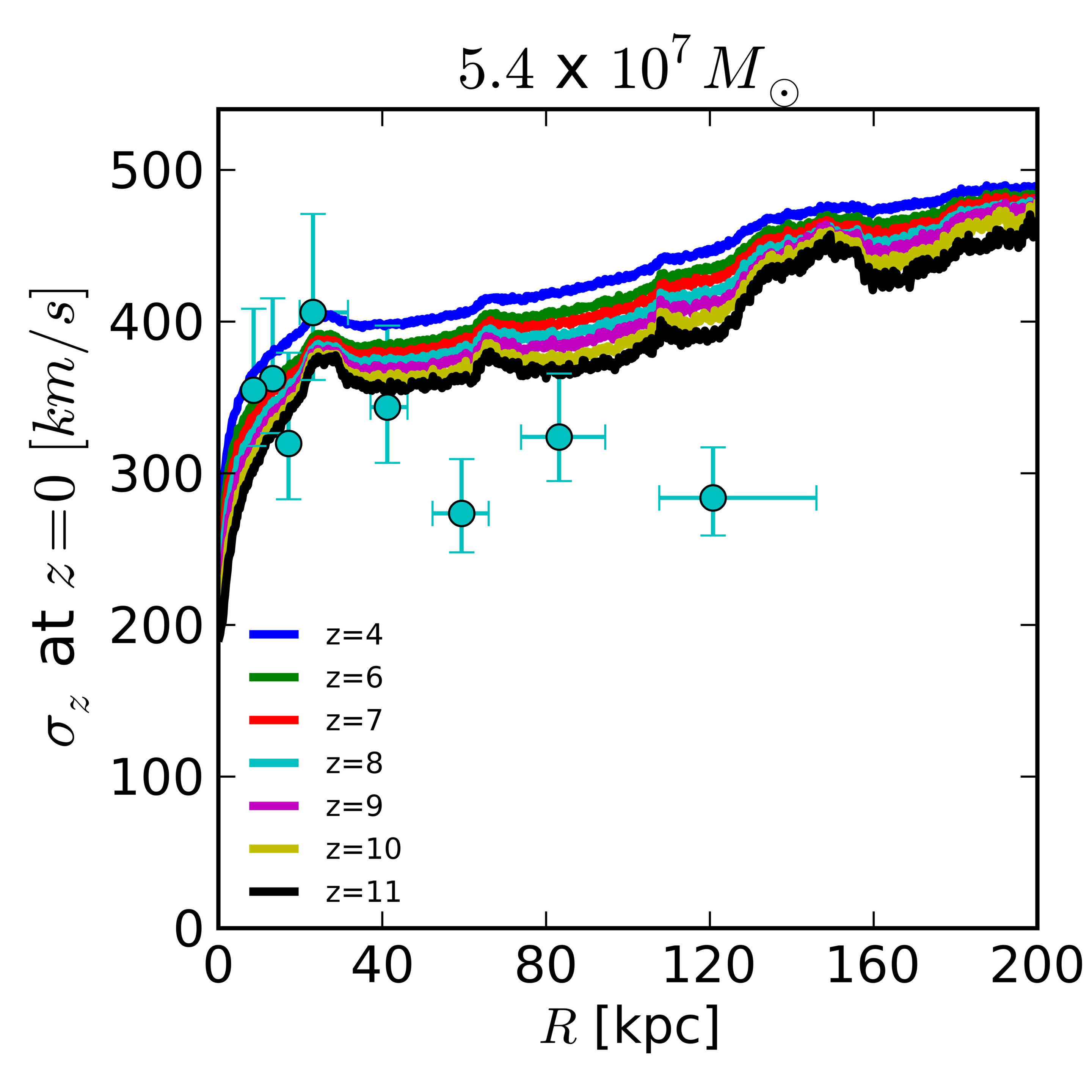}}
  \end{subfigure}
\begin{subfigure}{.95\linewidth}
  \centering
  \includegraphics[width=.95\linewidth]{{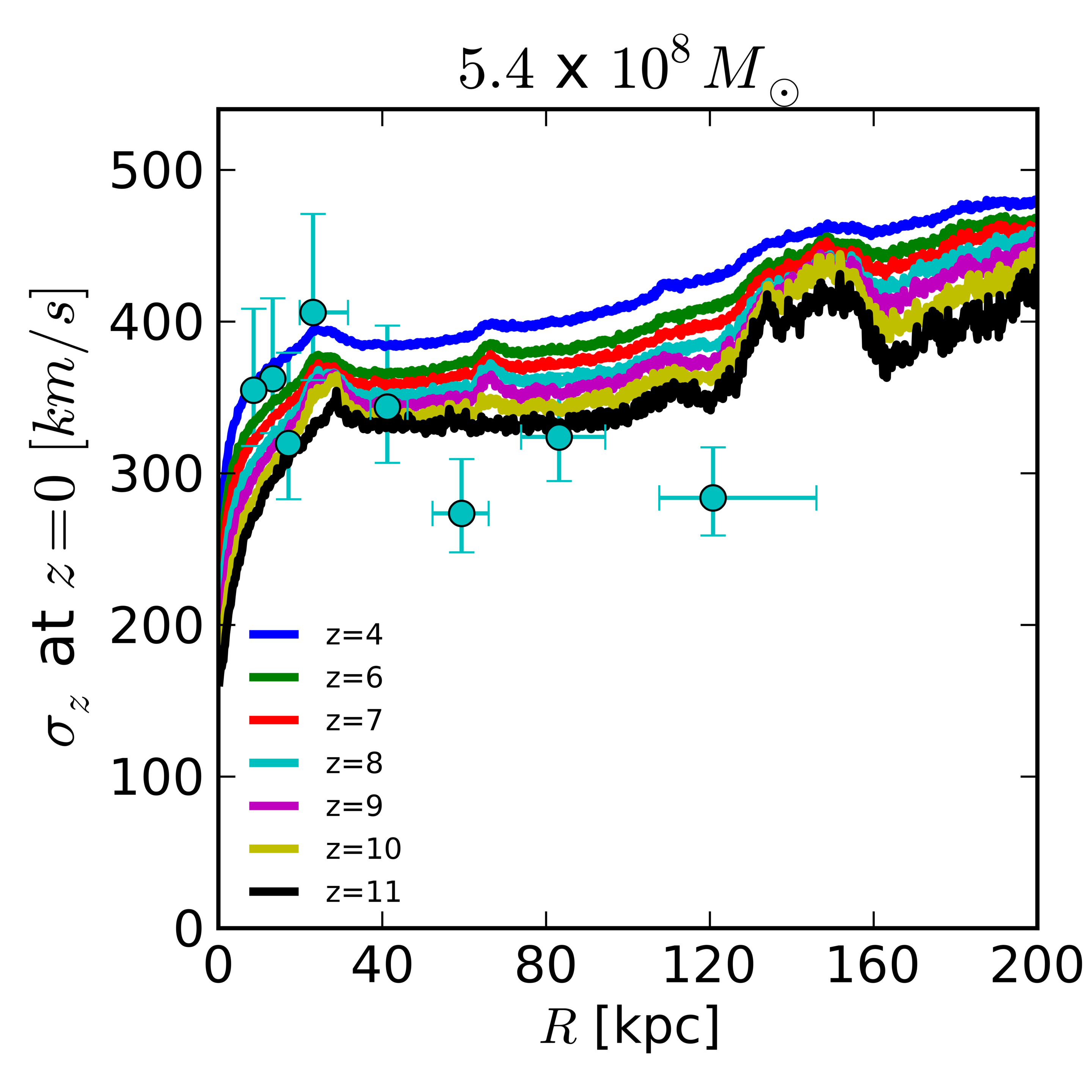}}
\end{subfigure}
\begin{subfigure}{.95\linewidth}
  \centering
  \includegraphics[width=.95\linewidth]{{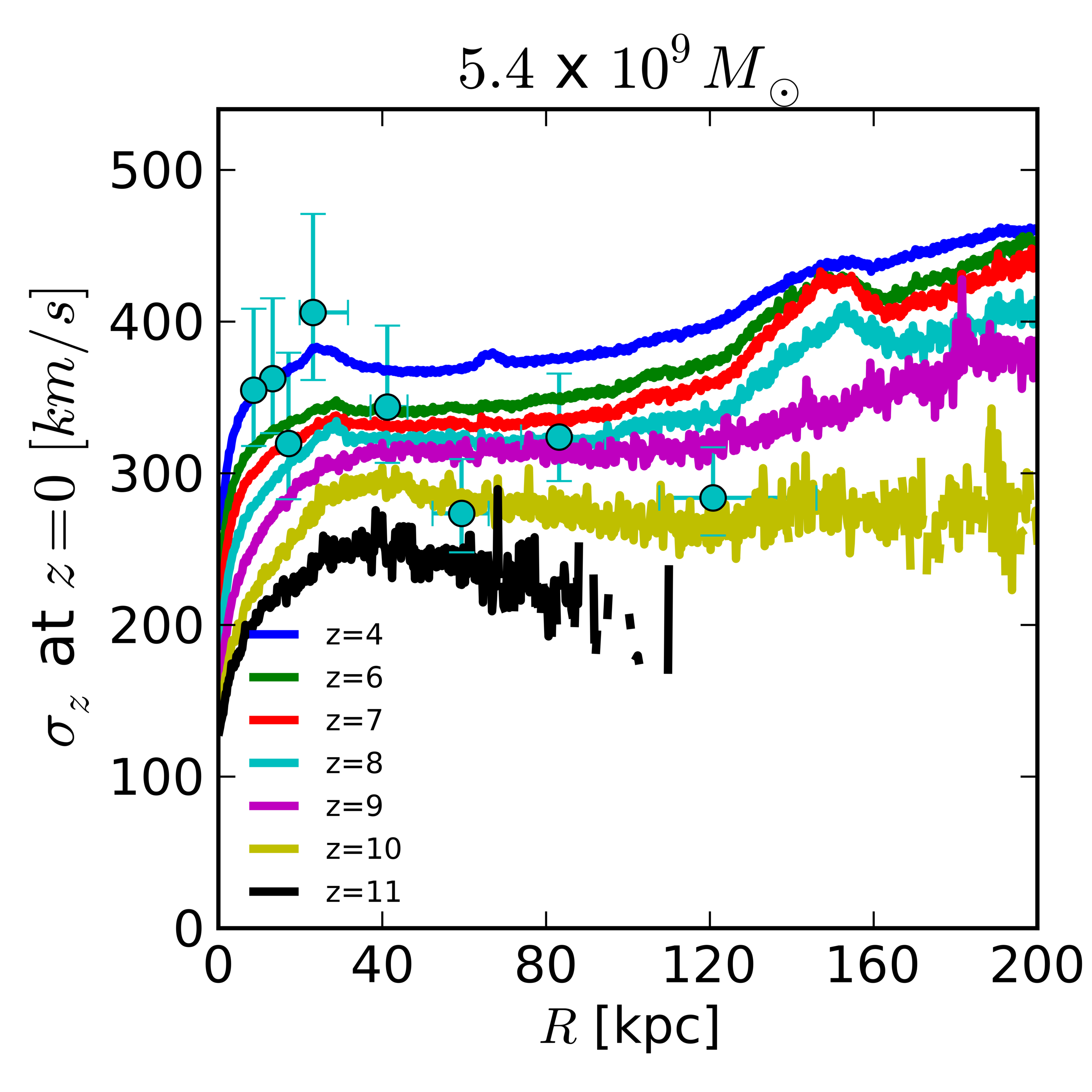}}
\end{subfigure}
\caption{Variation of the line of sight velocity dispersion with redshift cut for a fixed mass cut. Cyan points, with associated error bars, are observational data for M87 MPGCs \protect\citep{2011ApJS..197...33S}.}
\label{szfixedm}
\end{figure}

The variation of the line of sight velocity dispersion with redshift cut for fixed mass, as depicted in \textbf{Figure \ref{szfixedm}} shows that at $M\simeq 5 \times 10^7 \rm{M}_\odot$ all redshifts are significantly too hot, with $z\simeq11$ being the best fit. Again at $M\simeq 5 \times 10^8 \rm{M}_\odot$ the best fit is $z\simeq11$ with a degeneracy of redshifts, getting hotter as the redshift decreases but in general hovering around the same value and following the same trend. Finally for $M\simeq 5 \times 10^9 \rm{M}_\odot$ we have some discrimination between the velocity dispersions of various redshift cuts, and a good indication that $z\simeq9$ or $z\simeq10$ could be potential best fits.

The central region in the observational data shows an increase while our simulation results shows a sharp decrease. This is probably due to the absence of a BCG in our simulation results, which would cause the central increase in velocity dispersion seen in the observational data. Thus comparisons between simulations and observations are done outside of this region. Another caveat is that our halo mass is not perfectly normalized to that of M87, meaning that the general hotter-to-colder trend in the velocity dispersion would hold, no matter the normalization, but the exact best fitting value could change. 

The data show a downward trend  in velocity dispersion, and yet the simulations show in general an upward trend with radius. In general, higher sigma peaks produce colder distributions in the simulations and a flatter, though in general still rising, measurement of the dispersion outskirts of the halo. \cite{2005MNRAS.364..367D} see a similar effect when employing the Diemand-Moore technique; with increasing peak height the globular cluster distribution becomes both colder at $z=0$ and the velocity dispersion profile becomes shallower. Observationally, indications of a rising velocity dispersion at the edge of a galaxy when the ICL and halo stars are considered exist for some galaxies considered. In one of the 13 BCG galaxies considered in \cite{Fisher:1995ju} there is a stark example of a rising dispersion curve in BCG IC 1101 the dominant member of Abell 2029, previously observed to have a rising velocity dispersion in \cite{Dressler:1979jm}. Several additional BCGs analyzed in \cite{Fisher:1995ju} show a flat velocity dispersion profile. Our results predict the GC population of M87 should show a rising velocity dispersion curve. There is not yet a published velocity dispersion profile that we are aware of for M87 beyond $\sim 120\rm{kpc}$, but such data should offer confirmation to our analysis and provide an additional self consistent way to break the minimum mass/minimum redshift degeneracy in peak height of formation sites of globular clusters in M87. Recent results by \cite{Murphy:2014wb} show a hint of confirming our predictions, measuring a rising GC velocity dispersion in M87 out to 45kpc, and additional observational data beyond this will provide important constraints to our theoretical picture.

\subsubsection{Velocity Dispersion}
\begin{figure}
\begin{subfigure}{.95\linewidth}
  \centering
  \includegraphics[width=.95\linewidth]{{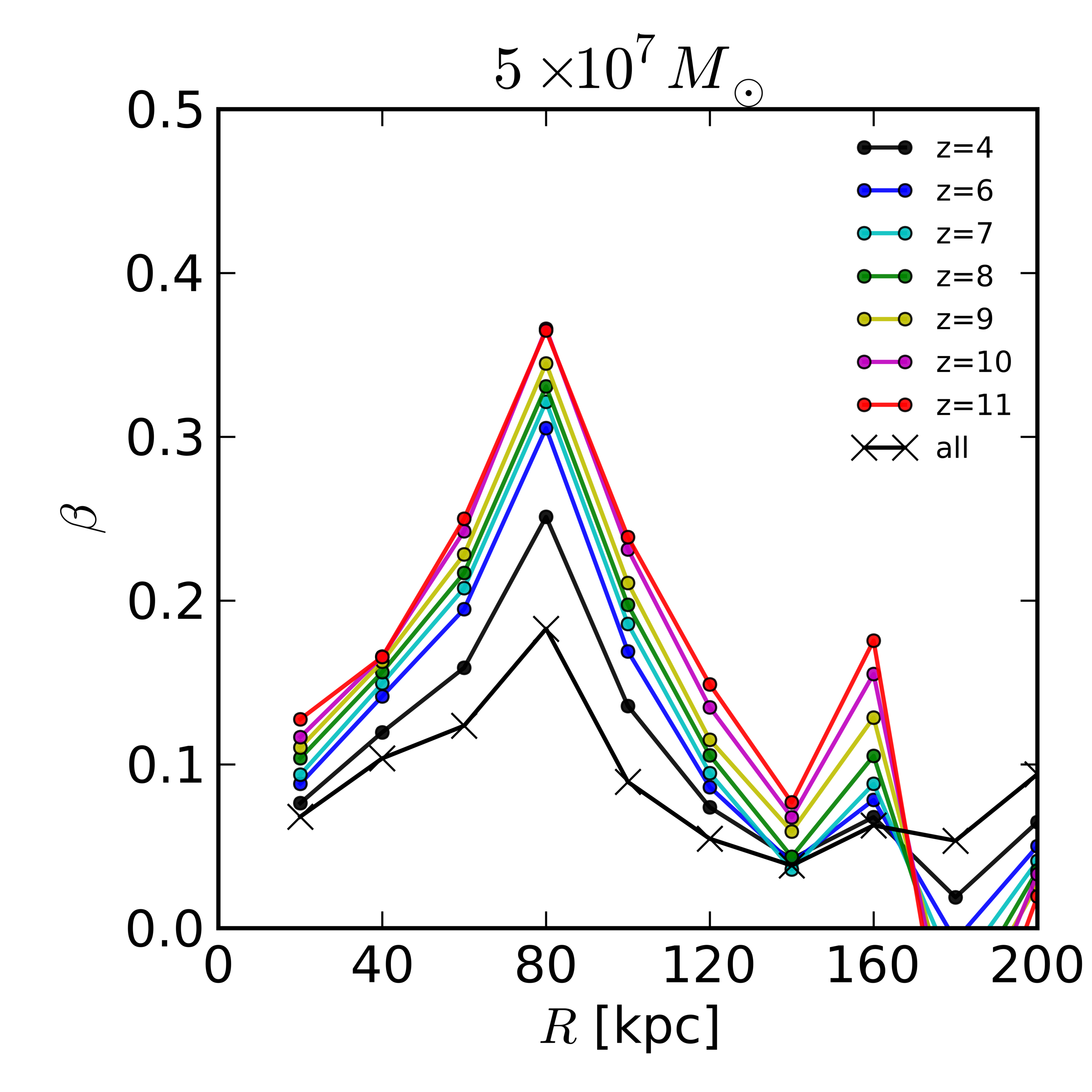}}
  \end{subfigure}
\begin{subfigure}{.95\linewidth}
  \centering
  \includegraphics[width=.95\linewidth]{{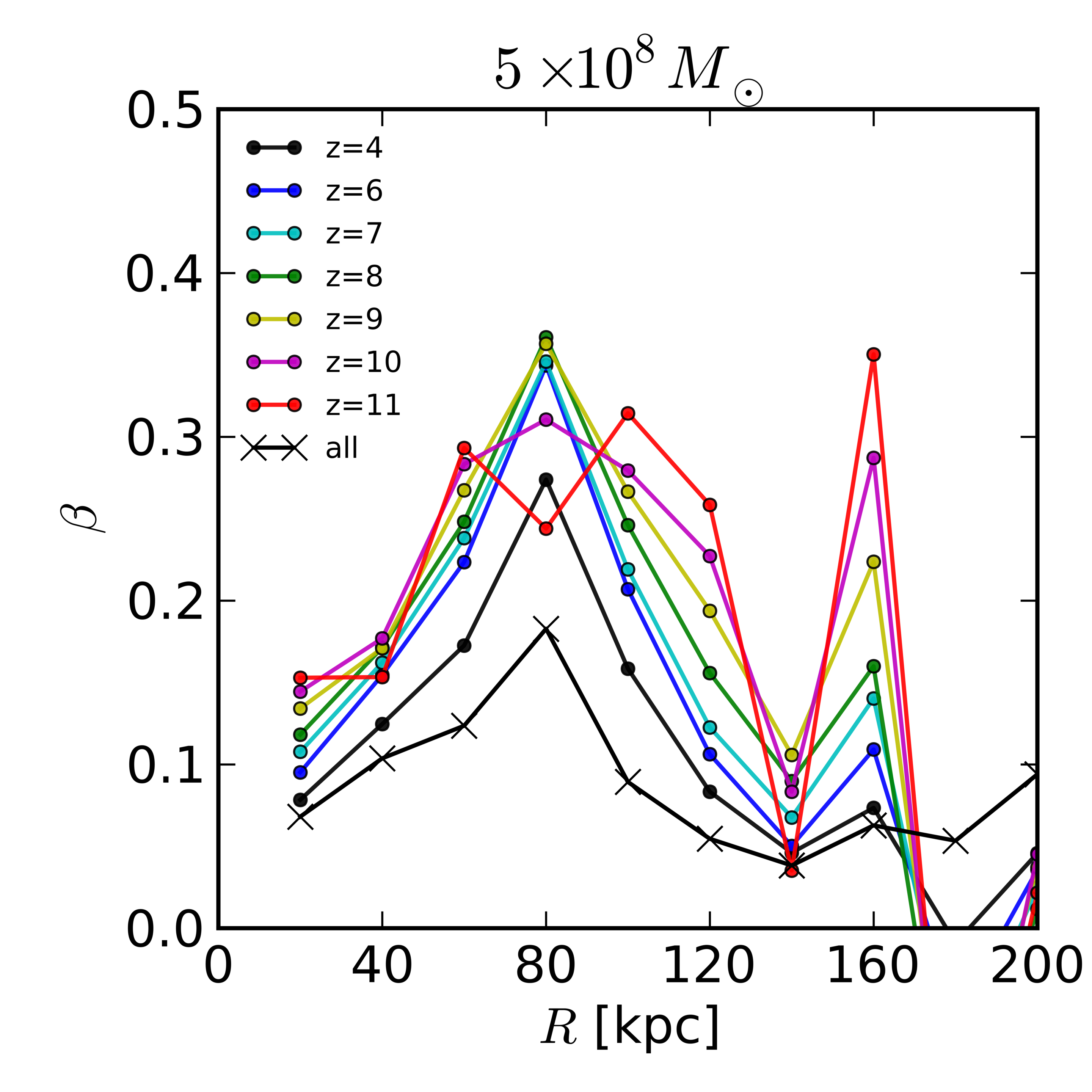}}
\end{subfigure}
\begin{subfigure}{.95\linewidth}
  \centering
  \includegraphics[width=.95\linewidth]{{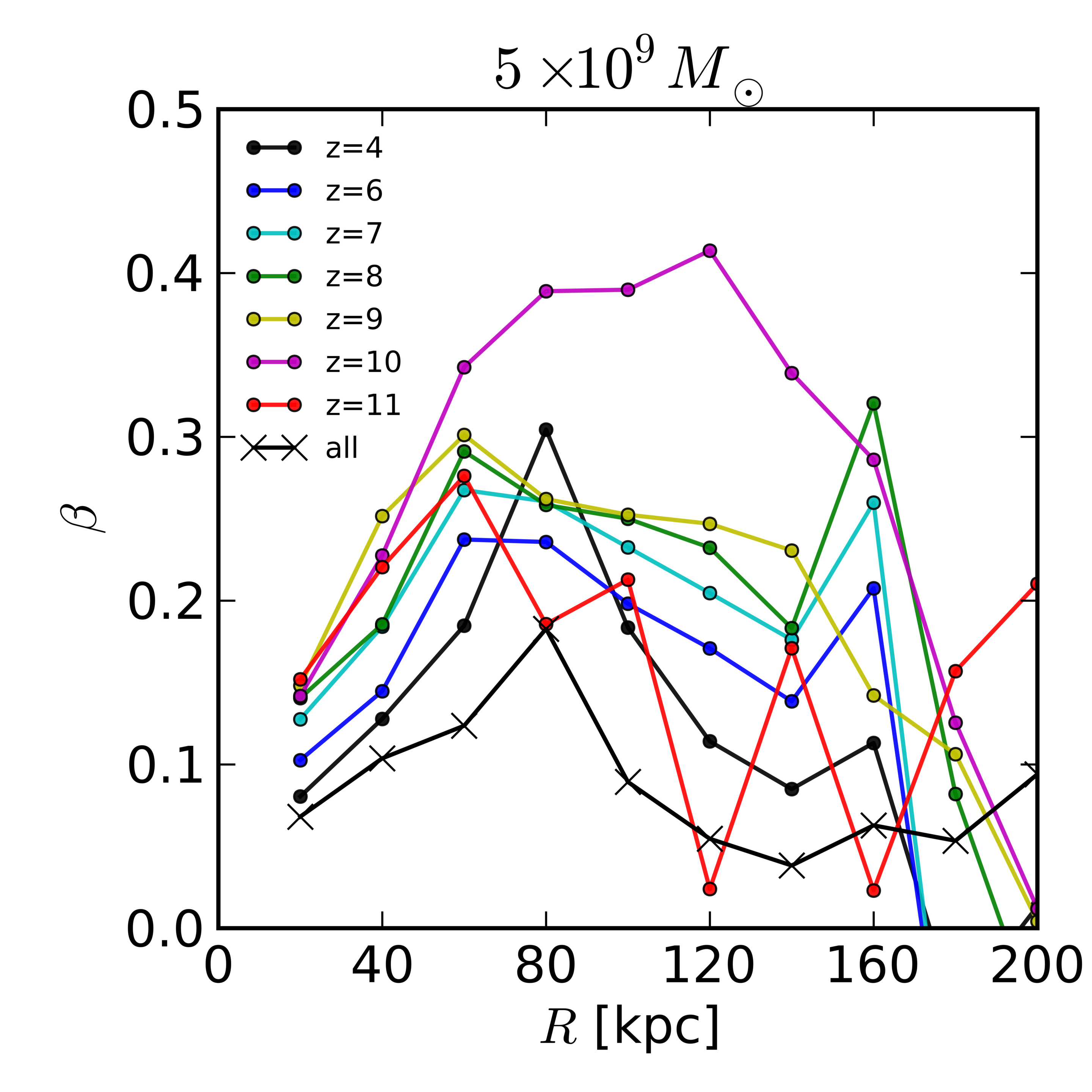}}
\end{subfigure}
\caption{$\beta$ for various sigma peaks at redshift 0}
\label{betato200kpc}
\end{figure}

The anisotropy parameter $\beta$ is defined as
\begin{equation}
\beta = 1 - \frac{1}{2}\frac{\sigma_t^2}{\sigma_r^2}
\end{equation}
and discriminates between radial and tangentially biased orbits. As orbits approach perfectly radial, $\beta \rightarrow 1$, $\beta <0$ which means the orbit is tangentially biased, and a perfectly circular orbit is perfectly tangentially biased i.e. $\beta \rightarrow -\infty$. $\beta$ is in general determined by the dependence of the distribution function on the total angular momentum \citep{2008gady.book.....B}.  Performing an analysis of $\beta$ on our halo, subtracting out the halo bulk velocity, we see that material originating in higher sigma peaks is more radially biased at $z=0$ as depicted in \textbf{Figure \ref{betato200kpc}}. As in \protect\cite{2012JCAP...10..049S} we see a radial bias in $\beta$ as a whole.

In terms of $\beta$ the Jeans equation for a spherical system is

\begin{equation}
\frac{d(\nu \bar{v_r^2})}{dr}+2\frac{\beta}{r} \nu \bar{v_r^2}= -\nu \frac{d \Phi}{dr}
\end{equation}
where $\nu$ is the density of the population, $\bar{v}$ the mean velocity.

Considering $\bar{v_r^2} \propto \sigma_z^2$ and $\nu \propto r^{-\alpha}$ we can rewrite the Jeans equation as
\begin{equation}
-\frac{\alpha}{r} \sigma_z^2 + \frac{d}{dr} \sigma_z^2 + \frac{2\beta}{r} \sigma_z^2 = - \frac{d \Phi}{dr}
\end{equation}

Assuming $\frac{d}{dr} \sigma_z^2$ is negligible, we solve for $\sigma_z^2$
\begin{equation}
	\label{sigmaz2}
\sigma_z^2=\frac{r}{2\beta-\alpha} \left(-\frac{d\Phi}{dr}\right)
\end{equation}

\begin{figure}
	\begin{subfigure}{0.95\linewidth}
	\centering
	\includegraphics[width=.95\linewidth]{{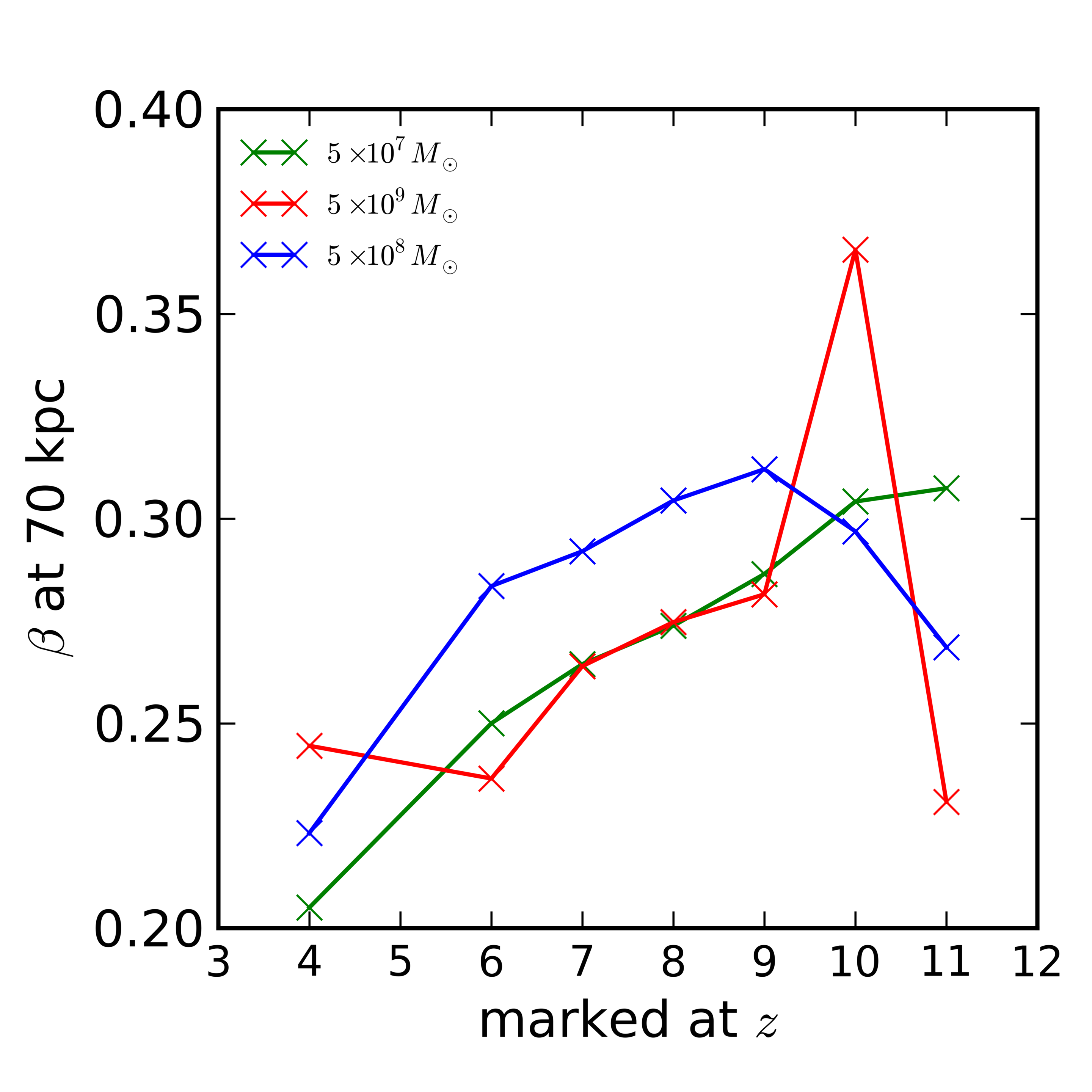}}
	\end{subfigure}
	\caption{$\beta$ as measured at 70 kpc}
	\label{betaaggregate}
\end{figure}

Material derived from a higher sigma peaks has a higher $\beta$ parameter as seen in \textbf{Figure \ref{betaaggregate}}, for such material from \textbf{Equation \ref{sigmaz2}} we expect the velocity dispersion to be in general lower. This is exactly the behavior we see in \textbf{Figure \ref{sigmazsynthesis}}, where for higher $z$ or higher mass, corresponding to material originating in higher sigma peaks, there is systematically a lower $\sigma_z$ with the effects being most prominent at the highest sigma level analyzed (e.g. $z\simeq11,m=5.4 \times 10^9 M_\odot$).

%

For two halos undergoing a merger, the centrally concentrated populations---exactly those originating from high-sigma peaks---behave more like point particles than two populations of more extended objects undergoing a merger.  We expect that after undergoing the merger the concentrated populations to maintain more of the memory of the merger event than the extended population, in which case the result is in effect smoothed. If the $\beta$ value is thus a function of an initial impact parameter of a merger event $b$ we would expect $\beta$ to be a stronger function of $b$ for centrally concentrated material than for the general population with the more centrally concentrated the material the stronger the effect. We see this clearly in \textbf{Figure \ref{betato200kpc}} where material originating in high sigma peaks follows the same trend as the general population simply exaggerated in effect the magnitude of the exaggeration increasing with increasing sigma. In this manner $\beta$ encodes information about past major mergers.
%
%

A few caveats to our results. Firstly, results in the literature generally smooth over hundreds of halos. \cite{2012JCAP...10..049S} quote a $\beta$ of $0.4$ for cluster sized halos independent of radius which was computed by averaging the effects of particular substructure or merger events through this process. We are dealing with a single sample, and so these effects are not averaged out. Secondly, \cite{2012JCAP...10..049S} show that $\beta$ is a simplistic analysis for non-spherical systems, as clusters such as in our analysis in general are, and propose a $\beta$ dependent on velocity dispersions along and perpendicular to the halo major axis. We thus consider our results as more of an indication of a trend than a proof of one.

\cite{Cote:2001bl} found that the MPGCs of M87 had a slight tangential bias of $\beta=-0.4$. We see this behavior as well simply at a larger radius with $\beta$ dipping below zero outside of 200 kpc before settling around 0 at the virial radius. This result concretely shows different dynamical behavior of the MPGCs subpopulation as we predict. The difference in our prediction amounts only to a difference in our merger history as simulated vs. the merger history as experienced by M87. While our simulation is an M87 analogue, it is naturally just one realization and its history is not demanded to be identical. The observations of \cite{Cote:2001bl} provide an important clue that our model of MPGC formation results in correct dynamical behavior at $z=0$.


\subsection{Star Formation Efficiency}

As a consistency check, we compare the mass we must produce in stars of globular clusters at a given mass and redshift cut (corresponding to a peak rarity $\nu$) to the mass that could have come from star formation at the marked redshift. Using a virial radius of $R\simeq1 $~Mpc \citep{2011MNRAS.414.2101U} and the mass model for Virgo from \cite{McLaughlin:1999ih}, we get a virial mass $M_{200}$ for Virgo of $\simeq 2-3 \times 10^{14} \rm{M}_\odot$. We note that our simulated cluster has a lower mass of $\simeq 1 \times 10^{14} \rm{M}_\odot$.

To estimate the present day BGC population mass in Virgo, we use the fits computed in \cite{Cote:2001bl} to the BGC population. Noting that the data extends to only $100$~kpc, we compute a lower and upper bound to the present day BGC population mass by extending our integration to $100$~kpc and $\simeq1$~Mpc. The integrated BGC mass we find are $\simeq 3 \times 10^9$ and $\simeq 4 \times 10^{10}$ respectively. Theoretical work which follows our hypothesis of MPGC formation agrees with our finding of a steep surface density profile of the distribution at $z=0$ \citep{Moore:2006bi,Spitler:2012fr,Boley:2009gf}, whereas the profile of \cite{Cote:2001bl} is designed to follow the gas distribution where data is unavailable resulting in a much shallower profile than predicted by our formation theory. Thus, to be consistent with our picture we regard the integrated mass past $100~\rm{kpc}$ as an overestimate when using the profile developed by \cite{Cote:2001bl}. 

To extend this to a BGC population mass at high redshift we account for GC destruction which reduces the mass function by a factor of 1.5-4  \citep{Boley:2009gf}. Using a fiducial value of 2, we get an integrated BGC population mass of $\simeq 5.6 \times 10^9-7.6 \times 10^{10} \rm{M}_\odot$ at high redshift. Finally, in order to compute the total baryonic mass at high redshift, we assume a realistic star formation efficiency in the range of $1\%-10\%$.

\begin{figure}
  \centering
  \caption{Ability to form the requisite number of stars by the given redshift for a (M,z) cut, assuming instantaneous truncation of reionization. The curves are the conditional mass function for a halo of  $3 \times 10^{14} \rm{M}_\odot$ at $z=0$. The gray region corresponds to the WMAP5 measurement for reionization. The magenta region corresponds to the best fitting model from this work, $z\simeq 9,M\simeq 10^8 \rm{M}_\odot$ with a width equal to the difference in $\nu$ between two our best fitting models. The black lines correspond to the baryonic mass that is required to produce the present day BGC population.}
  \label{boley}
  \includegraphics[width=.95\linewidth]{{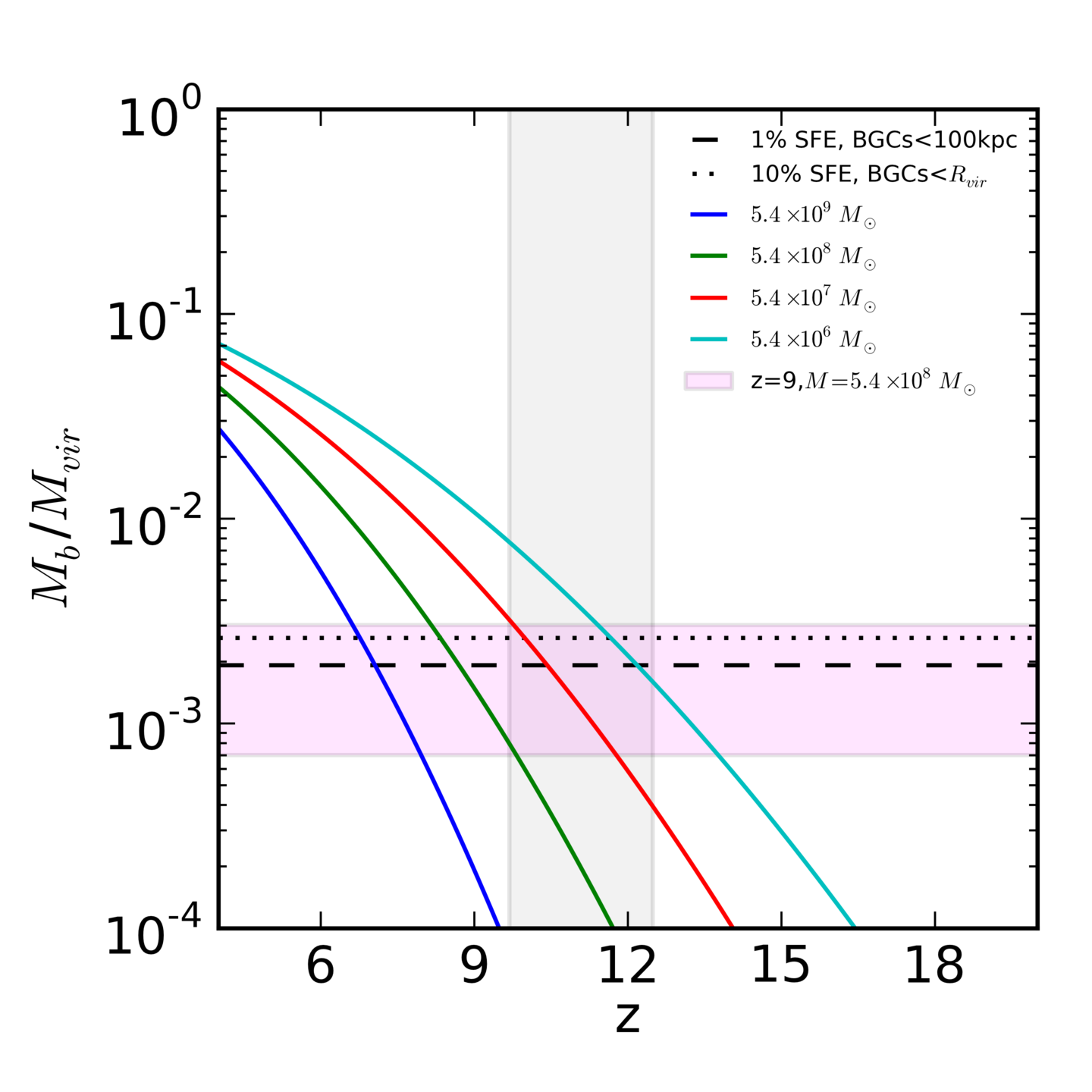}}
\end{figure}

Assuming a WMAP5 cosmology, we compute the baryon fraction in collapsed objects for a given dark matter halo minimum mass cutoff by computing the conditional mass function using the Sheth and Tormen \citep{Sheth:2002fn} formalism for a halo of $3 \times 10^{14} \rm{M}_\odot$ at $z=0$. \textbf{Figure \ref{boley}} depicts this result. The gray region corresponds to the WMAP5 measurement for reionization, and the magenta region corresponds to the best fitting model from this work. We note that the uncertainty of the population mass can be absorbed by the uncertainty in the star formation efficiency, and that a model with the population mass upper bound and 1\% star formation efficiency is equivalent to the lower bound and 10\% star formation efficiency, and that both are consistent with our best fitting model of $z\simeq9,M\simeq 10^8 \rm{M}_\odot$. 

 We can see that for the mass curves to intersect with the dashed lines of baryonic mass necessary to be generated that $M=10^9 \rm{M}_\odot$ is basically ruled out, but for very high redshift of reionization outside of the typical window. We see that lower mass cuts are consistent with lower overdensity peaks (meaning lower redshift as well) and correspond to lower star formation efficiency. Intermediate mass cuts $10^7-10^8 \rm{M}_\odot$ are consistent with higher sigma peaks requiring high star formation efficiency to be consistent with observations. These results provide a consistency check and show that with a realistic star formation efficiency assumption we can consistently form the requisite BGCs by the redshift of truncation, which is itself consistent with independent reionization constraints.

\section{Resulting Constraints on Reionization}
\label{sec:synthesis}
To summarize the constraints, we performed a linear  fit in ($log(R)$,$log(\Sigma(R))$ space on the surface density profile, with slope $m$, so as to not unduly weight the central points, where error bars come from the variation of that fit over our 100 random lines of sight. This was then compared to the fit to the observational data used by \cite{Spitler:2012fr}. For the velocity dispersion profile, we compared the results at $70~\rm{kpc}$ to the observational data used by \cite{Spitler:2012fr}.

\begin{figure}
  \centering
  \caption{Linear fit in log/log to the density profile at $z=0$, at each mass threshold marking criterion and marking redshift simulated. }
\label{sigmadsynthesis}
  \includegraphics[width=.95\linewidth]{{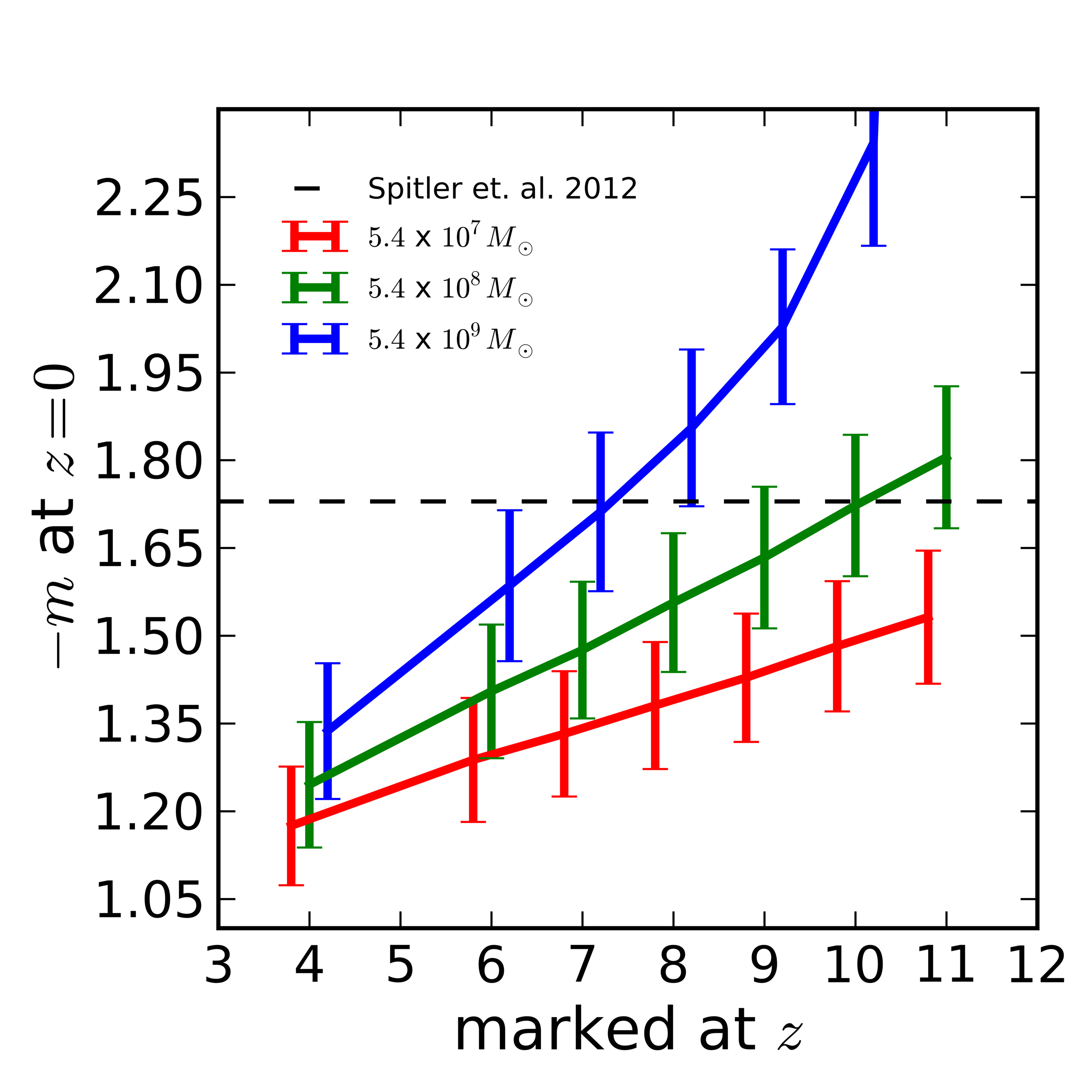}}
\end{figure}

\begin{figure}
  \centering
  \caption{the $\sigma_z$ value at $70$~kpc at each mass threshold marking criterion and marking redshift simulated. }
\label{sigmazsynthesis}
  \includegraphics[width=.95\linewidth]{{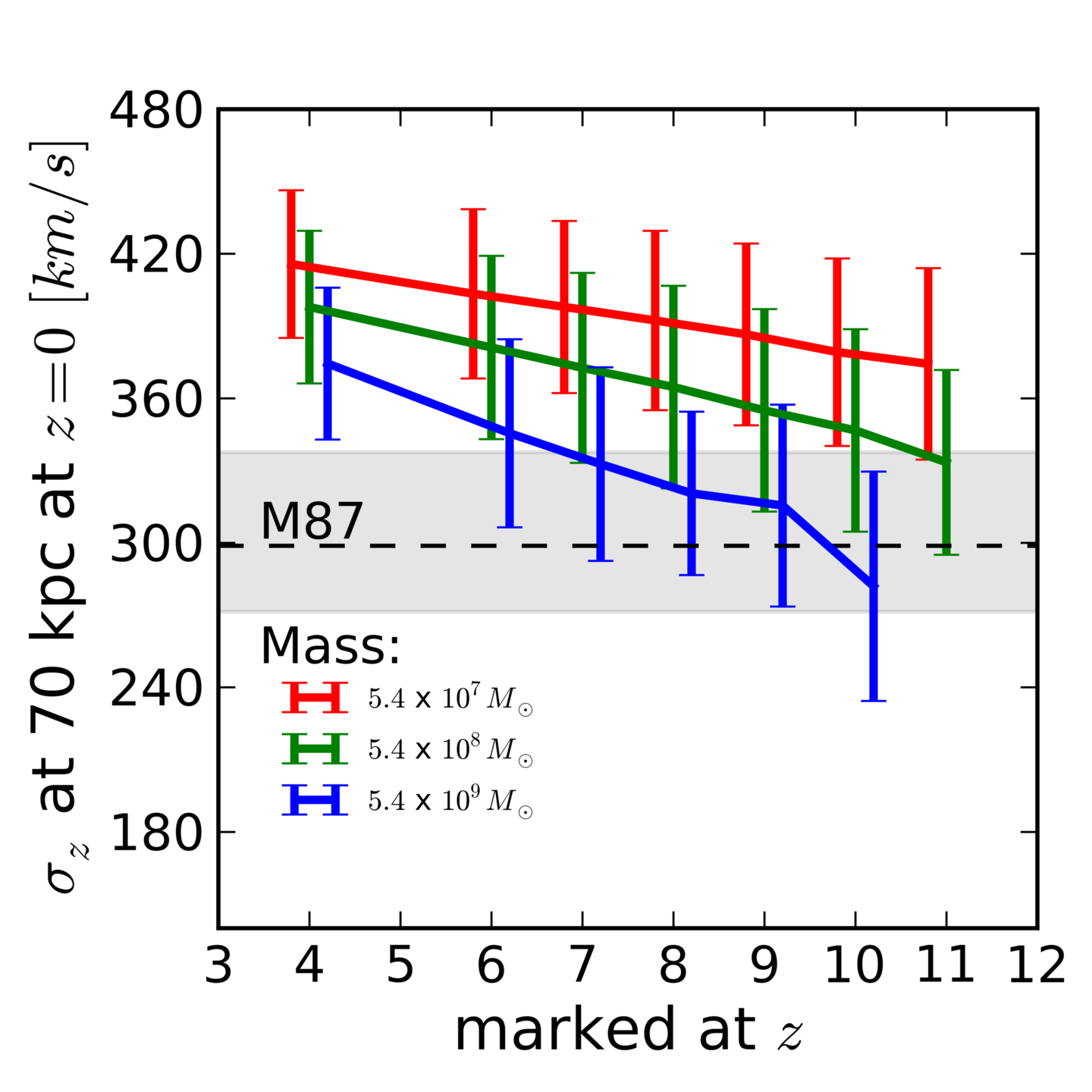}}
\end{figure}

\textbf{Figure \ref{sigmadsynthesis}} and \textbf{Figure \ref{sigmazsynthesis}} depict these summary results, respectively and show that for the line of sight velocity dispersion $z\simeq 10$ is the best match for $\simeq 5 \times 10^9 \rm{M}_\odot$ and $z\simeq 11$ is the best match for $\simeq 5 \times 10^8 \rm{M}_\odot$ while $\simeq 5 \times 10^7 \rm{M}_\odot$ points to a cut at much higher redshift outside of our simulation output range. For the surface density profiles, the slope of the linear fit consistently points to the fact that the higher the mass cut, the lower the redshift cut, with similar values of $z\simeq 7$ for $\simeq 5 \times 10^9 \rm{M}_\odot$, $z\simeq 9-10$ for $\simeq 5 \times 10^8 \rm{M}_\odot$ and $\simeq 5 \times 10^7 \rm{M}_\odot$ points to much higher than $z\simeq 12$.

As done in \cite{Spitler:2012fr} we are able to put constraints on reionization in a Virgo-like object. We use the Diemand-Moore technique to constrain the rarity of the halos that metal poor globular clusters form within. If the hypothesis that metal poor globular clusters formed in high sigma peaks and that their formation was indeed truncated by reionization is correct our joint constraints from the velocity dispersion, surface density distribution, and star formation efficiency analysis point to a mass of $\sim5.4 \times 10^8 \rm{M}_\odot$ ( $\simeq 5 \times 10^9 \rm{M}_\odot$ is ruled out by star formation efficiency analysis, and  $\simeq 5 \times 10^7 \rm{M}_\odot$ ruled out unless reionization completed before $z\simeq 11$). With a mass of $\sim5.4 \times 10^8 \rm{M}_\odot$ reionization there is some tension, with the velocity dispersion being a slightly better fit at higher redshift cuts, but with the velocity dispersion not being an especially tight constraint. Thus we can consistently defer to the more discriminating surface density distribution profile at this mass cut, which clearly points to a value of $z \simeq 9$

\begin{figure}
  \centering
  \caption{Surface density for $z\simeq 9,\simeq 5 \times 10^8 \rm{M}_\odot$. For reference we include the model most similar to the best fitting model of Spitler et al. (2012) $z\simeq 6,\simeq 5 \times 10^8 \rm{M}_\odot$. Both are normalized to the observational value in at $70$ kpc.  Plotted are the averages over all 100 lines of sight analyzed with error bars indicating the dispersion of the average. The black line is the analytic fit arrived on by \protect\cite{Spitler:2012fr}, and we see that it fits the observational data equally well as our best fitting model. Cyan points, with associated error bars, are observational data for M87 MPGCs \protect\citep{2011ApJS..197...33S}.}
\label{sigmadbestfit}
  \includegraphics[width=.95\linewidth]{{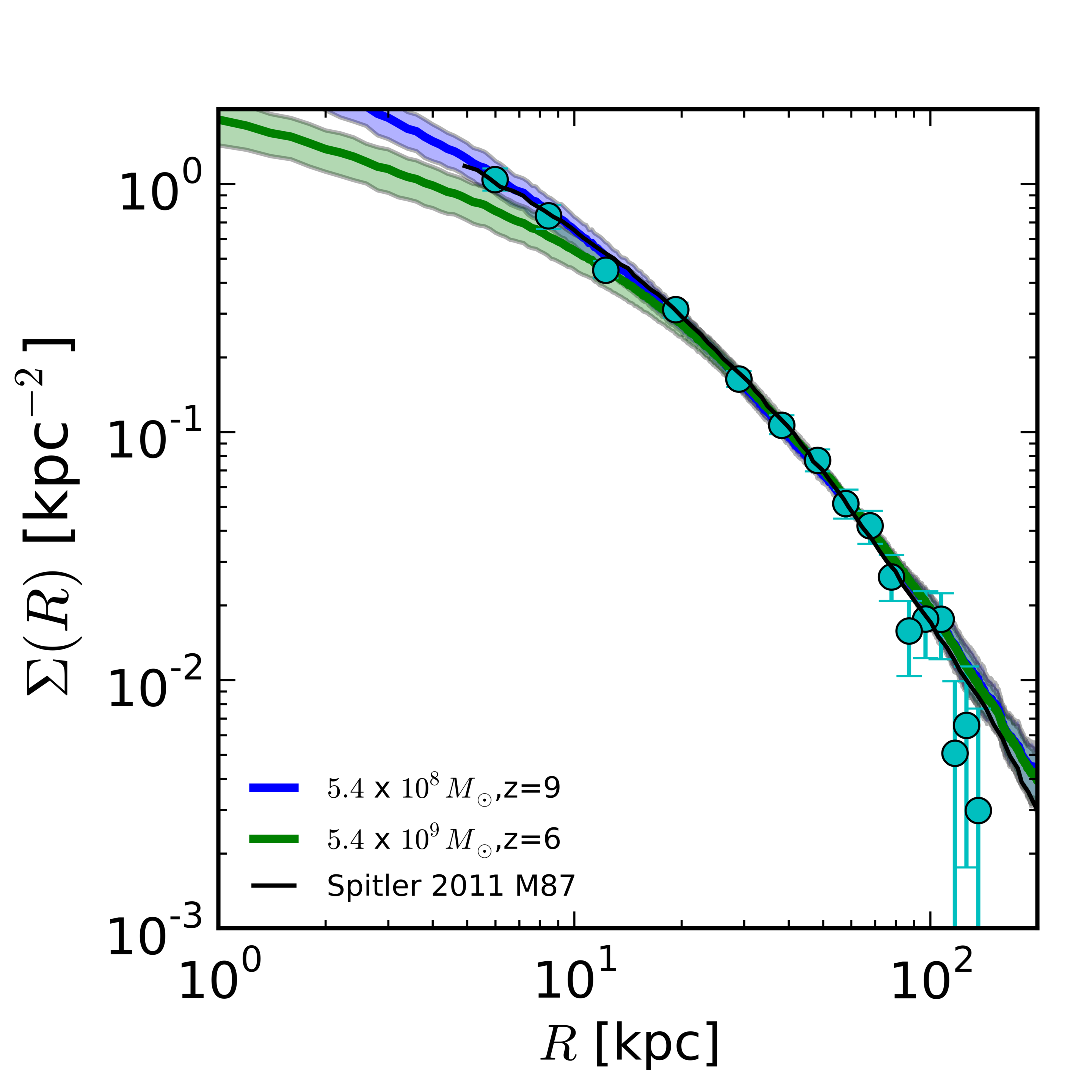}}
\end{figure}

\begin{figure}
  \centering
  \caption{Line of sight velocity dispersion for $z\simeq 9,\simeq 5 \times 10^8 \rm{M}_\odot$. For reference we include the model most similar the best fitting model of \protect\cite{Spitler:2012fr} $z\simeq 6,\simeq 5 \times 10^8 \rm{M}_\odot$. Both are normalized to the observational value in at $70$ kpc. Plotted are the averages over all 100 lines of sight analyzed with error bars indicating the dispersion of the average. The black line is the analytic fit arrived on by \protect\cite{Spitler:2012fr}, we can see that while it correctly accounts for the BCG which our dark matter only simulation does not, and thus is a better fit at inner radii, it performs poorly at greater radii and does not predict the rising velocity dispersion with radius, as we do. Cyan points, with associated error bars, are observational data for M87 MPGCs \protect\citep{2011ApJS..197...33S}.}
  \includegraphics[width=.95\linewidth]{{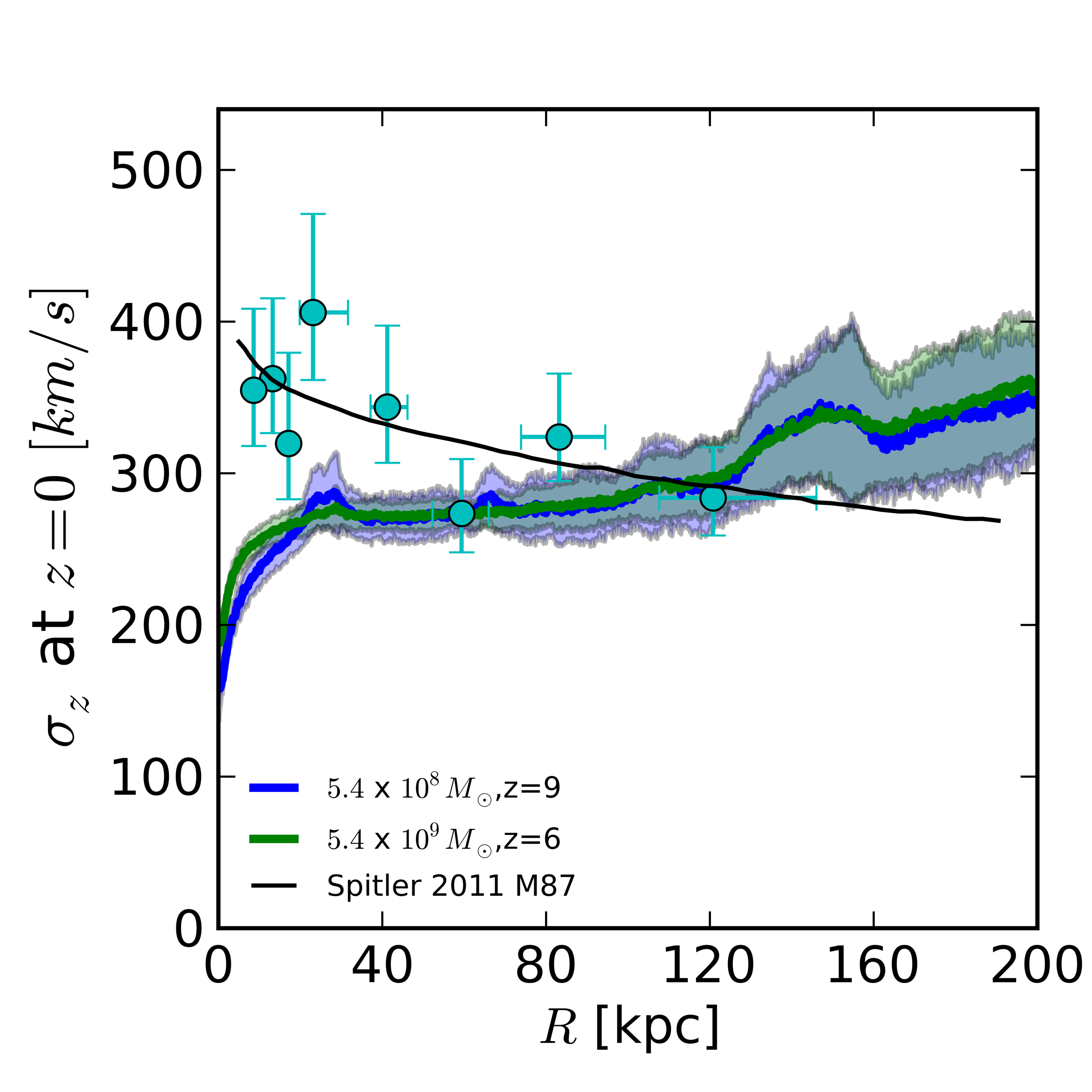}}
\label{sigmazbestfit}
\end{figure}

We thus present for our best fitting model $z\simeq 9,M=5.4 \times 10^8 \rm{M}_\odot$ the surface density and line of sight velocity. Surface density distribution results for the best fitting model are shown in \textbf{Figure \ref{sigmadbestfit}} and line of sight velocity dispersion results for the best fitting model are shown in \textbf{Figure \ref{sigmazbestfit}}. Plotted are the averages over all 100 lines of sight analyzed with error bars indicating the dispersion of the average. For reference we include the model most similar to the best fitting model of \cite{Spitler:2012fr} $z\simeq 6,M= 5.4\times 10^8 \rm{M}_\odot$

\section{Intracluster Globular Clusters}

Simulations of cluster galaxies suggest that between 10\% and 50\% of stars residing in a cluster are intracluster, that is unattached to any galaxy \citep{Murante:728140,2004MNRAS.355..159W,2005MNRAS.357..478S}, these simulations predict that the intracluster population is older and more centrally concentrated. These can include globular clusters themselves. Virgo stands out as the nearest system to have a significant intracluster stellar population \citep{2004IAUS..217...88C}.

Intracluster globular clusters (IGCs) are an important probe of the history of galaxies and clusters \citep{1995ApJ...453L..77W} and can constrain the systems dynamical history. However, obtaining a large sample of these objects is observationally difficult due to their faint nature and the need to pick them out against the background of large galaxies. An observational study of Virgo's IGCs found 4 metal poor IGC  candidates \citep{2007ApJ...654..835W}, if this sample representative of the population as a whole the majority of IGCs are expected to likewise be metal poor a prediction which was corroborated by \citep{2010Sci...328..334L}. Our theoretical work can shed light on this population and estimate the frequency IGCs in a Virgo like cluster informing further observational searches seeking to exploit IGCs as cosmological probes.

To identify a population of IGCs, we used the structure and substructure phase space finder \texttt{ROCKSTAR} \citep{2013ApJ...762..109B} to identify halos and subhalos and within our simulation. \texttt{ROCKSTAR} is a 6D, optionally 7D (temporal), phase-space halo finder which first divides the volume into 3D Friends-of-Friends groups  using a large linking length as a parameter. An adaptive phase-space metric is then built so that for each group, 70\% of its particles are linked in subgroups, a procedure which is recursively applied. The final step assembles the seed halos in their densest subgroups assigning each particle to the group closest to it in phase space and performs an unbinding procedure to remove unbound particles. 

We deploy \texttt{ROCKSTAR} in its 6D implementation to form a full halo catalog at $z=0$. This catalog provides positions and radii, among other halo properties, for each halo and subhalo. We use this information to define the intracluster globular cluster population as a subset of the tracer population of our best fitting model. Particles residing within a halo identified by \texttt{ROCKSTAR} are considered to trace the bound distribution for observational purposes and those which do not are considered to trace the IGC distribution. 

\begin{figure}
  \centering
  \caption{Intracluster globular clusters. Intracluster is defined as not residing within any structure or substructure found by the halo finder \texttt{ROCKSTAR}. In our favored model within $46~\rm{kpc}$ we have a total mass of $7.2 \times 10^{11} M_{\odot}$ in particles we identify as tracing the globular cluster distribution. We use the ratio $7.2 \times 10^{11} M_{\odot}$/161 as our conversion factor to normalize the number distribution of IGCs we obtain.}
  \includegraphics[width=.95\linewidth]{{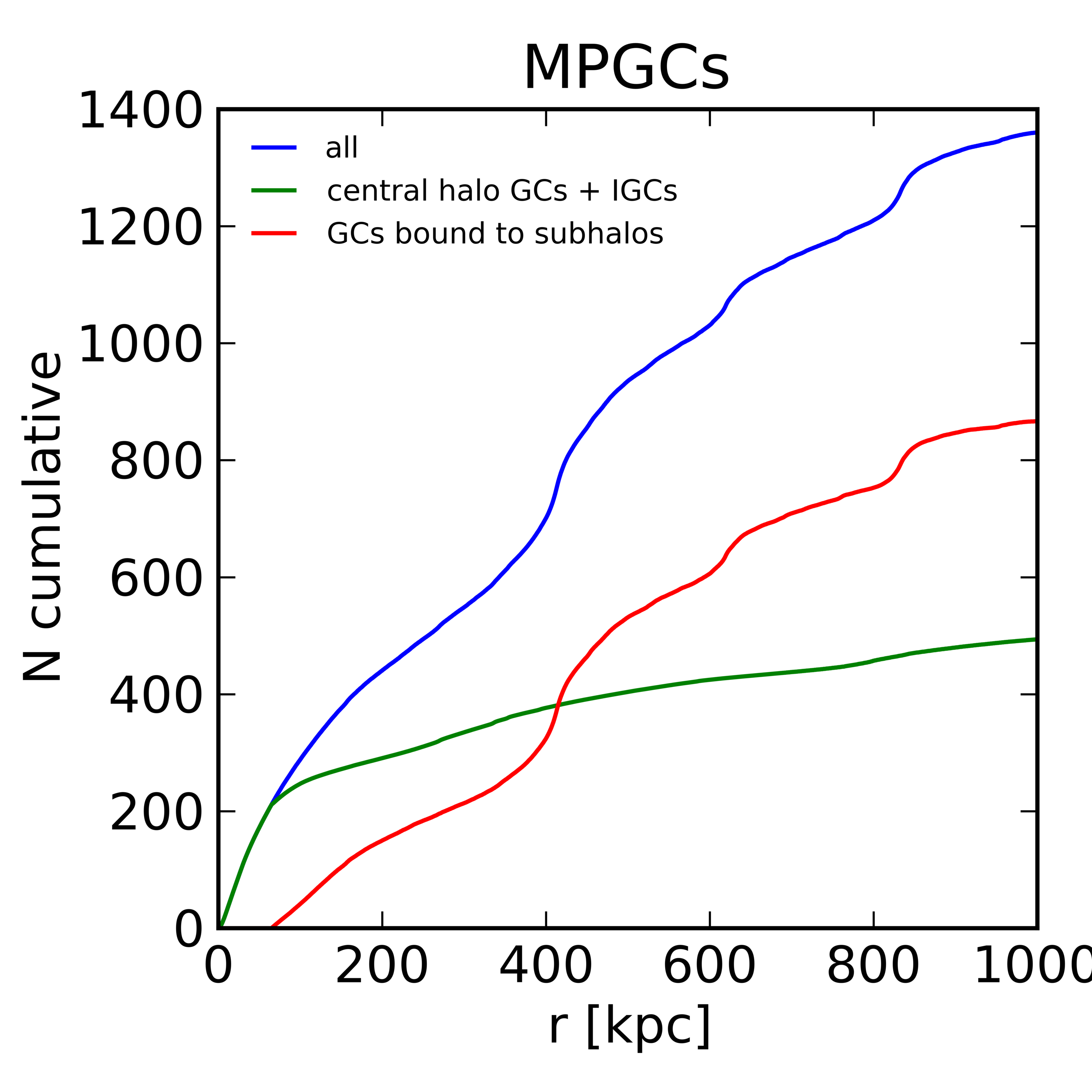}}
\label{freefloatingpop}
\end{figure}

To convert this population into an observational estimate we note that the M87 GC system by \cite{Cote:2001bl} includes data out to $46~\rm{kpc}$ and comprises a robust detection of 278 GCs, of which 161 are considered blue metal poor GCs. In our favored model within $46~\rm{kpc}$ we have a total mass of $7.2 \times 10^{11} M_{\odot}$ in particles we identify as tracing the globular cluster distribution. We use the ratio $7.2 \times 10^{11} M_{\odot}$/161 as our conversion factor to normalize the number distribution of IGCs we obtain. About 20\% of our total GC population is defined as intracluster in  our analysis. This is on the same order of the estimate of previous analyses by \cite{2005MNRAS.364L..86Y} who find that about 30 \% of all GCs in a rich cluster are intracluster and that IGCs are likely to have a flatter profile, which we likewise see. We can see that the GCs indeed tend to be centrally concentrated around structure as if the analysis were performed for general material a full 43\% would be defined as intracluster. The cumulative fraction of IGCs vs. radius is given in \textbf{Figure \ref{freefloatingpop}}, we predict $\simeq 300$ IGCs at the virial radius. 


\begin{figure}
  \centering
  \caption{Bound globular clusters. Bound is defined as residing within any structure or substructure found by the halo finder \texttt{ROCKSTAR}. In our favored model within $46~\rm{kpc}$ we have a total mass of $7.2 \times 10^{11} M_{\odot}$ in particles we identify as tracing the globular cluster distribution. We use the ratio $7.2 \times 10^{11} M_{\odot}$/161 as our conversion factor to normalize the number distribution of IGCs we obtain.}.
  \includegraphics[width=.95\linewidth]{{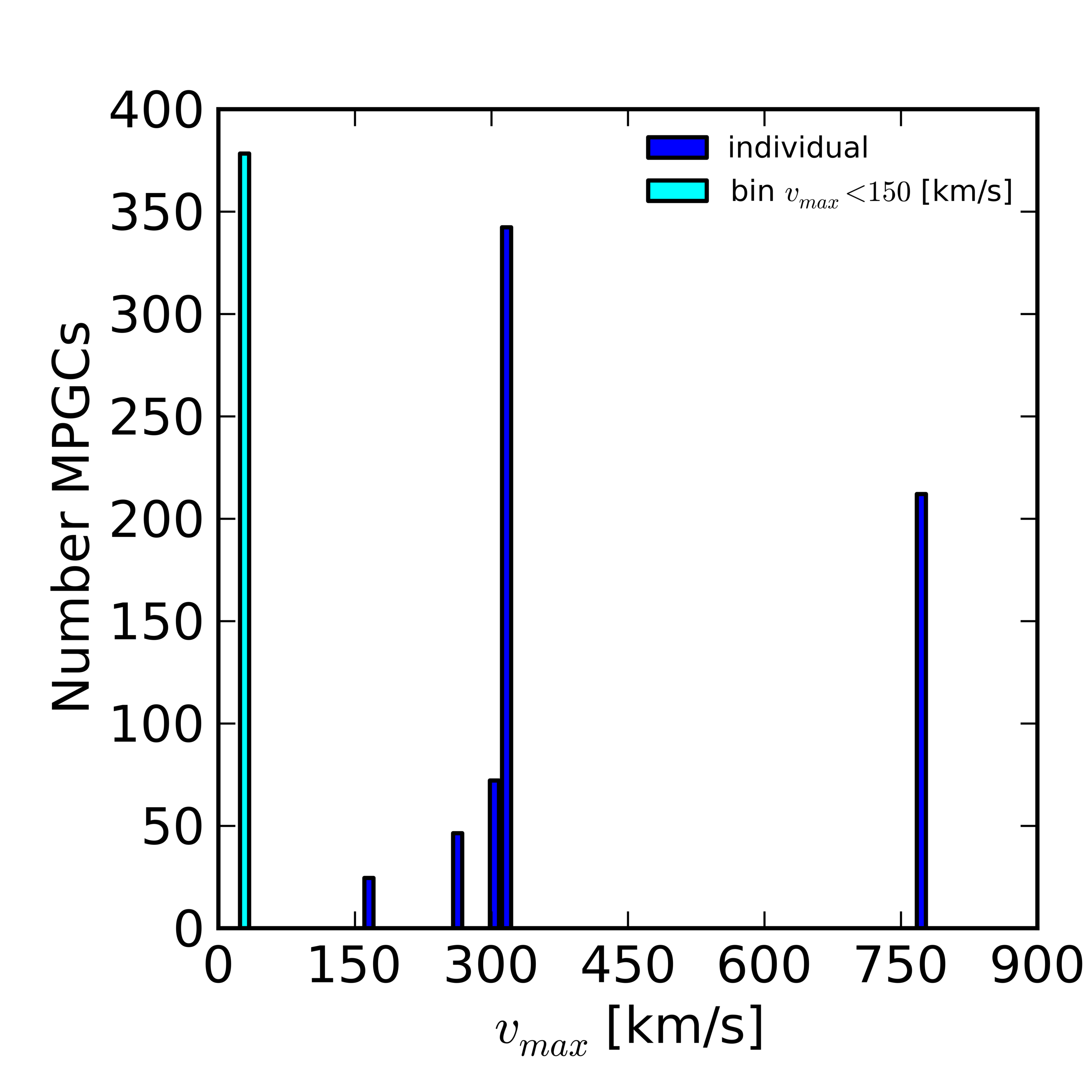}}
\label{boundpop}
\end{figure}

To complete our analysis, we examine bound MPGCs as a function of their $v_{\rm{max}}$ noting that this value is for the entire halo, so is in general much higher than that computed for the galaxy alone. We normalize the MPGC distribution such that the largest halo contains 161 MPGCs. We see  in \textbf{Figure \ref{boundpop}} that the number of MPGCs in smaller halos  $v_{\rm{max}}<150$  is negligible compared to both the total number of bound MPGCs and the total number of intracluster MPGCs, and in general a smaller halo of $v_{\rm{max}}<150$  is expected to host $<1$ MPGC on average. 

\section{Discussion}
Our best fitting model disagrees with \cite{Spitler:2012fr} who found M87 reionized later at $z\sim6$. However, we clearly see from the comparison plots that the models are degenerate and both fit the data within the error bars. The best fitting model therefore must be chosen by incorporating additional information to break this degeneracy. We can explain the difference in our conclusion by refinements in the more precise constraints were are able to achieve through a numerical procedure with the relaxed assumption of mass model and spherical symmetry as well as to our choice as to how to address the $M,z$ degeneracy. Specifically \cite{Spitler:2012fr} use an additional  mass threshold based on halos cooling via atomic cooling only to select star forming halos, corresponding to halos with a virial temperature above $\rm{T}_{\rm{vir}}=10^4~\rm{K}$ With a similar assumption, we get similar results. However, we prefer to relax this assumption, allowing for the possibility of molecular cooling \citep{Tegmark:1997ce,2002Sci...295...93A}, and choose the address the $M,z$ degeneracy instead through a star formation efficiency argument in \textbf{Figure \ref{boley}} as done in \cite{Boley:2009gf}.

We see that the line of sight does indeed have a significant impact on the line of sight velocity dispersion observed. Interesting to note is that as the dependency on the surface density depends only logarithmically on line of sight, there is little scatter from the mean in the surface density as seen in summary in \textbf{Figure \ref{sigmadsynthesis}} and plotted for one $M,z$ pair in \textbf{Figure \ref{sigmadbestfit}}, whereas the line of sight velocity dispersion is quite sensitive to the line of sight chosen and can vary by up to$\sim100~\rm{km/s}$ in our results as seen in summary in \textbf{Figure \ref{sigmazsynthesis}} and plotted for one $\nu$ in \textbf{Figure \ref{sigmazbestfit}}. We see that error bars on these quantities are consistent across redshift and mass pairs.

A point for discussion is in our model there is a tension between the velocity dispersion data and the surface density distribution results, this tension is balanced by our best fitting model. Considering the surface density distribution results, these point to high mass, intermediate redshift formation, or low mass high redshift formation sites and epochs, respectively. Whereas considering the velocity dispersion results alone, they point to a high redshift and higher mass halo formation sites i.e. very large $\nu$; these are the only pairs which produce a flat velocity dispersion curve which seems to correspond to the data and analytic fits of \cite{Spitler:2012fr}. This is an interesting hint that if MPGCs indeed form in a $\nu$ larger than our best fitting results, that in the extended in time formation model picture that at low redshift MPGCs would be required to form in very high mass halos $\simeq 10^{11}$ and above to maintain the same peak rarity a conclusion that may be difficult to reconcile with data. 

\section{Conclusions} 
We build upon work by \cite{Spitler:2012fr} identifying MPGCs as forming in high sigma peaks of the density distribution at high redshift. We propose, validate, and utilize a novel split resolution simulation technique to push resolution to the requisite level to address to confirm the analytic results of \cite{Spitler:2012fr} numerically in an Virgo Cluster analogue. We quantify the dependency of a chosen line of sight for measurements of surface density and velocity dispersion of the globular cluster population in an Virgo Cluster analogue, finding that surface density measurements to be robust across all lines of sight, the velocity dispersion numerical measurements could have up to a $\sim 100$~km/s spread. We find the dependency on the chosen line of sight at $z=0$ of the globular cluster population to be consistent across density peak heights of the initial formation sites. Our results are:
\begin{itemize}
\item We show that our favored model $z~\simeq~9,M\simeq 5 \times 10^8 \rm{M}_\odot$ can consistently form the requisite BGCs by $z\simeq9$, a redshift consistent with reionization constraints.
\item Our best fitting model relaxes \cite{Spitler:2012fr}'s restricted assumptions of an analytic model employing spherical symmetry and atomic cooling only. A redshift of $z\simeq 9$ is more consistent with observational evidence for the reionization window in the cluster environment, vs. the much lower $z\simeq 6$ picked out with the Spitler et al. technique.
\item The tension in the velocity dispersion points naturally to further constraints that could come from observations past 130~kpc in M87. We predict a rising velocity dispersion profile. The mass normalization and concentration parameter difference between our simulated cluster and M87 could both shift the velocity dispersion profile upward, or even potentially influence its slope. 
\item We predict $\simeq 300$ intracluster MPGCs in the Virgo cluster. Better observational constraints on the number density of intracluster MPGCs, particularly at high radii, would support or falsify our formation scenario.
\item Baryonic physics, realistic globular cluster evaporation modeling, modeling tidal stripping would bring the velocity dispersion simulation results further in line with observations, particularly at the center of the halo. 
\end{itemize}
\section{Acknowledgements}
Simulations were performed on the Monte Rosa system at the Swiss Supercomputing Center (CSCS). C.C.M. was supported by the HP2C program through the Swiss National Science Foundation.

\bibliographystyle{mn2e}
\bibliography{GCsMNRAS}

\bsp

\label{lastpage}

\end{document}